\documentclass[a4paper,fleqn,usenatbib,useAMS]{mnras}

\usepackage{graphicx}	
\usepackage{amsmath}	
\usepackage{amssymb}	
\usepackage{multicol}        
\usepackage{bm}		
\usepackage{pdflscape}	
\usepackage[T1]{fontenc}
\usepackage{aecompl}

\newcommand{\kms}{\,km\,s$^{-1}$} 
\def\oii{[{\sc O\,II}]}

\def\nii{[{\sc N\,II}]}

\def\g09{G10}
\def\logmass{$\log(M_*)$}
\def\ms{$\log(M_*/M_\odot)$}
\def\ha{{\rm H$\alpha$}}
\def\aha{$A_{H\alpha}$}

\def\kms{km s$^{-1}$}
\def\tdhst{3D-\textit{HST}}
\def\recal{\textsc{Recal-L025N0752}}

\def\lsim{\mathrel{\hbox{\rlap{\hbox{\lower4pt\hbox{$\sim$}}}\hbox{$<$}}}}
\def\gsim{\mathrel{\hbox{\rlap{\hbox{\lower4pt\hbox{$\sim$}}}\hbox{$>$}}}}

\def\sn{$(S/N)_{\rm peak}$}

\def\z{\textit{z}}
\def\Mo{\cal{M}_{\odot}}
\def\Msunyr{M$_{\odot}$ yr$^{-1}$}

\usepackage[T1]{fontenc}
\usepackage{ae,aecompl}

\usepackage{mathptmx}

\title[ROLES H$\alpha$]{\textit{HST }\ha~grism spectroscopy of ROLES: a flatter low-mass slope for the z$\sim$1 SSFR--mass relation}

\author[R.~Ramraj et al.]
  {Riona Ramraj$^{1,2}$, David G.~Gilbank$^1$\thanks{E-mail: gilbank@saao.ac.za}, Sarah-Louise Blyth$^2$, Rosalind E. Skelton$^1$
  \newauthor
  Karl Glazebrook$^3$,  Richard G.~Bower$^4$ and Michael L.~Balogh$^{5}$ 
\\
$^1$South African Astronomical Observatory, P.O Box 9, Observatory, 7935, Cape Town, South Africa\\
  $^2$Department of Astronomy, University of Cape Town, Private Bag X3, Rondebosch 7701, South Africa\\
  $^3$Centre for Astrophysics and Supercomputing, Swinburne University of Technology, P.O. Box 218, Hawthorn, VIC 3122, Australia\\
$^4$Institute for Computational Cosmology, Department of Physics, University of Durham, South Road, Durham, DH1 3LE, UK\\
 $^5$Department of Physics and Astronomy, University of Waterloo, Waterloo, Ontario, Canada N2L 3G1\\}

\date{Last updated 2015 May 22; in original form 2013 September 5}

\pubyear{2015}

\begin{document}
\label{firstpage}
\pagerange{\pageref{firstpage}--\pageref{lastpage}}
\maketitle

\begin{abstract}
We present measurements of star formation rates (SFRs) for dwarf galaxies (M$_{*}$$\sim$10$^{8.5}$$\Mo$) at z$\sim$1 using near-infrared slitless spectroscopy from the \textit{Hubble Space Telescope (HST)} by targetting and measuring the luminosity of the \ha~emission line. Our sample is derived from the Redshift One LDSS3 Emission Line Survey (ROLES), which used \oii$_{\lambda3727}$ as a tracer of star formation to target very low stellar masses down to very low SFRs ($\sim$0.1 $\Mo$yr$^{-1}$) at this epoch. 
 Dust corrections are estimated using SED-fits and we find, by comparison with other studies using Balmer decrement dust corrections, that we require a smaller ratio between the gas phase and stellar extinction than the nominal Calzetti relation, in agreement with recent findings by other studies. 
By stacking the WFC3 spectra at the redshifts obtained from ground-based \oii~detections, we are able to push the WFC3 spectra to much lower SFRs and obtain the most complete spectroscopic measurement of the low mass end of the SSFR--mass relation to date. We measure a flatter low mass power law slope (-0.47 $\pm$0.04) than found by other (shallower) \ha-selected samples ($\approx$-1), although still somewhat steeper than that predicted by the EAGLE simulation (-0.14 $\pm$0.05), hinting at possible missing physics not modelled by EAGLE or remaining incompleteness for our \ha~data.
\end{abstract}

\begin{keywords}
galaxies: dwarf --
galaxies: evolution --
galaxies: general
\end{keywords}



\section{Introduction}
\label{sec:introduction}
There have been many surveys studying the star formation properties of  statistical samples of galaxies out to high redshifts (e.g. \citealt{1996ApJ...460L...1L}, \citealt{1996MNRAS.283.1388M}, \citealt{2006ApJ...651..142H}, \citealt{2011ApJ...730...61K}, \citealt{2012A&A...539A..31C}). However, most of these surveys are biased towards the most massive or most actively star forming galaxies. In order to better understand how galaxies form and evolve, we also need to study the low mass (dwarf) galaxies. Dwarf galaxies are more numerous than their higher mass counterparts and are considered the building blocks of high mass galaxies in the hierarchical formation scenario (e.g. \citealt{1978MNRAS.183..341W}, \citealt{1991ApJ...379...52W}, \citealt{2006Natur.440.1137S}). 

Studies have shown that the cosmic star formation rate (SFR) has declined by an order of magnitude since \z$\sim$1, when the Universe was approximately half its current age (e.g. \citealt{2006ApJ...651..142H}). This period represents the end of the peak of SFR activity in the Universe \citep{2014ARA&A..52..415M} which is frequently dubbed `Cosmic Noon'. By studying galaxies at this epoch and combining with studies at the present epoch, we can determine how galaxies have evolved over this time (i.e., the processes that have caused star formation to decline) and use this to constrain galaxy formation models (e.g. \citealt{2013MNRAS.431.3373H}, \citealt{2012MNRAS.422.2816B}).

The galaxy population can be divided broadly into red quiescent early type galaxies and blue star-forming late type galaxies (e.g. \citealt{2004ApJ...600..681B}, \citealt{2004ApJ...615L.101B}). The physical processes governing the transition from the blue cloud to the red sequence are not yet well understood. One key discovery of recent years is the tight sequence correlating stellar mass and SFR for star-forming galaxies (\citealt{2004ApJ...600..681B}, \citealt{2007ApJ...660L..43N}, \citealt{2010aMNRAS.405.2594G}, \citealt{2014ApJ...795..104W}), usually refered to as the ``main sequence" or ``star-forming sequence". This main sequence is seen out to \z~$\sim$1 (e.g. \citealt{2006ApJ...647..853W}, \citealt{2007ApJ...660L..43N}) and even \z~$\sim$2 (e.g. \citealt{2009ApJ...706L.173B}). The remarkable constancy of shape and smooth evolution of this relation argues for relatively smooth star-formation histories of star-forming galaxies. In conjunction with the relatively rapid growth of passive galaxies, this hints at a smooth replenishment of star-forming galaxies as they are ``quenched" \citep{2007ApJ...665..265F}. Of particular importance is the pushing of measurements to lower stellar masses to enable any curvature of the main sequence to be measured (e.g., \citealt{2012ApJ...754L..29W}) as this, combined with the evolution of the stellar mass function, is an important consistency check of our measurements and of galaxy evolution models \citep{2015ApJ...798..115L}. One of the vital ingredients which must be tested is the SFR indicator being used, which can vary between surveys and between redshifts within the same survey. 

SFR indicators such as H$\alpha$ and the UV emission essentially measure the ionizing flux from young, hot massive stars while the mid or far-infrared emission measures the amount of the ionizing flux absorbed and re-radiated by dust. Most of the flux emitted by these young stars is at UV wavelengths. Measuring the ionizing flux from nebular emission lines (e.g. H$\alpha$ recombination line and, indirectly, the \oii~forbidden line) in a galaxy's spectrum is one way of estimating the SFR of a galaxy.
In order to study evolution, one would ideally like to measure the SFR using a single indicator which can be applied from low to high redshifts. In practice, this is difficult because each indicator is subject to different biases and selection effects. However, H$\alpha$ is considered the most direct indicator because it traces the current star formation in a galaxy and is less affected by dust (e.g. \citealt{1998ARA&A..36..189K}; K98) than shorter wavelength emission lines (such as \oii) and has a small dependence on metallicity \citep{2001MNRAS.323..887C}.

At \z~$\leq$ 0.4, the \ha~line is typically used for studying evolution because it can be observed optically (e.g. \citealt{2007ApJ...657..738L}, \citealt{2010ApJ...712L.189D}). \ha~moves out of the optical range at \z~$>$ 0.4 into the NIR. For this reason, the \oii~emission line which is available in the optical out to \z$\sim$1.5 has been used instead (e.g. \citealt{2009ApJ...701...86Z}, \citealt{2010bMNRAS.405.2419G},  \citealt{2011MNRAS.413.2883B}).
Previously, NIR spectrographs on large telescopes enabled observations of only a few tens of galaxies at a time at z~$\sim$1 because these were restricted to longslit spectroscopy (e.g. \citealt{1999MNRAS.306..843G}, \citealt{2002MNRAS.337..369T}). Recently, the development of NIR multi-object spectrographs and wide-field NIR cameras with narrow band filters has enabled large ground-based \ha~surveys to be conducted (e.g. \citealt{2008MNRAS.388.1473G}, \citealt{2008ApJ...677..169V},  \citealt{2012MNRAS.420.1061T}, \citealt{2013AJ....145...47M},  \citealt{2013ApJ...777L...8K}) out to \z $\sim$ 2.5. It is difficult to deal with atmospheric effects, such as seeing which blurs the image and the sky brightness which adds background noise, from ground-based observations. Space-based telescopes provide a better alternative because they are above the atmosphere and therefore exclude these effects. In particular, the much lower background in the NIR from orbit makes space particularly advantageous when compared with ground based observations. \ha~spectroscopic surveys with $HST$ for example, provide much deeper observations than is possible from the ground (e.g. \citealt{1999ApJ...520..548M}, \citealt{2009ApJ...696..785S}). (With optical surveys, where we are able to work between bright sky lines, the difference between ground and space is much less pronounced.) The Wide-Field Camera (WFC3) and grism on \textit{HST} has been used to conduct slitless spectroscopic surveys, for example \tdhst~(\citealt{2012ApJS..200...13B}, \citealt{2011ApJ...743L..15V}) and  WISP \citep{2010ApJ...723..104A}.

In this paper, we study a spectroscopic sample of  \oii-selected dwarf galaxies by targetting the \ha~emission line using near-infrared (NIR) slitless spectroscopy from HST. We determine the \ha~SFR and compare it to the SFR derived from the \oii~fluxes, confirming that the mass-dependent correction found by \citet{2010aMNRAS.405.2594G} (G10, hereafter) is necessary to reconcile the SFR indicators for these $z\sim 1$ galaxies.

This paper is presented as follows:  Section 2 introduces the data used and explains how the reduction was done to extract the spectra for our sample of galaxies. A line detection algorithm was developed to analyze the extracted spectra and is described in \S3. The algorithm produces measurements of the line luminosity which we convert into \ha~SFR measurements and compare with other SFR indicators such as the mass-dependent correction for the \oii~SFR (\citealt{2010aMNRAS.405.2594G}; G10)  in \S4. We show the \ha~SSFR-mass relation for galaxies at low stellar masses ($M_{*}$$<$10$^{9.5}$$M_{\odot}$) in \S4, and \S5 presents our conclusions. All magnitudes are quoted on the AB system and we adopt a $\Lambda$CDM cosmology with $\Omega_{m}$=0.3, $\Lambda$=0.7 and \textit{H$_{0}$}=70 km s$^{-1}$ Mpc$^{-1}$. Throughout, we convert all quantities to those using a \cite{2001MNRAS.322..231K} initial mass function (IMF).

\section{Method}
\subsection{Sample Selection}
Our galaxies are selected from the Redshift One LDSS3 Emission line Survey (ROLES; \citealt{2009MNRAS.395L..76D}, \citealt{2010bMNRAS.405.2419G}). ROLES was designed to specifically target K-faint (22.5 $<$ $K_{AB}$ $<$ 24) star-forming dwarf galaxies at $\z\sim$1. ROLES obtained spectroscopy using a custom KG750 band-limiting filter on the LDSS3 spectrograph (on the Magellan II telescope) to cover the wavelength range 7500 $\pm$ 500 \textrm{\AA}. It targeted low stellar mass galaxies (8.5 $<$ log$(\frac{M_{*}}{M_{\odot}})$ $<$ 9.5) in the redshift range 0.89 $<$ $\z$ $<$ 1.15 in two fields: the Great Observatories Origins Deep Survey-South (GOODS-S) field and the FIRES field. In this paper we use only the GOODS-S data, which has overlapping WFC3 grism spectroscopy from the 3D-HST survey.

The \oii$_{\lambda3727}$ emission line was used to obtain spectroscopic redshifts and \oii~luminosities for estimating SFRs, down to a limit of $\sim$0.1 M$_{\odot}$yr$^{-1}$ \citep{2010bMNRAS.405.2419G}. The ROLES low mass data are supplemented with an external subsample of emission line galaxies from ESO public spectroscopy \citep{2008A&A...478...83V} to extend their mass range up to 10$^{11.5}$ M$_{\odot}$ to study the mass dependence of galaxy properties such as the SSFR-mass, star formation rate density (SFRD), luminosity function, etc., at \textit{z}$\sim$1. The ESO public spectroscopic data overlap the same region of sky as ROLES and only those galaxies within the ROLES redshift range were selected. The \oii~SFRs were measured in the same way as for ROLES by \cite{2010bMNRAS.405.2419G}.  To first order, the ESO public and ROLES are just different mass sub-samples of otherwise similar data. A more detailed comparison can be found in \cite{2010bMNRAS.405.2419G} and is also discussed in the appendix.
\\
The full data sample (i.e. ROLES and ESO galaxies) and the sample containing ROLES galaxies only are hereafter referred to as WFC3-OII and WFC3-ROLES respectively.

\subsection{Data}
\begin{figure*}
\centering
\includegraphics[width=18cm]{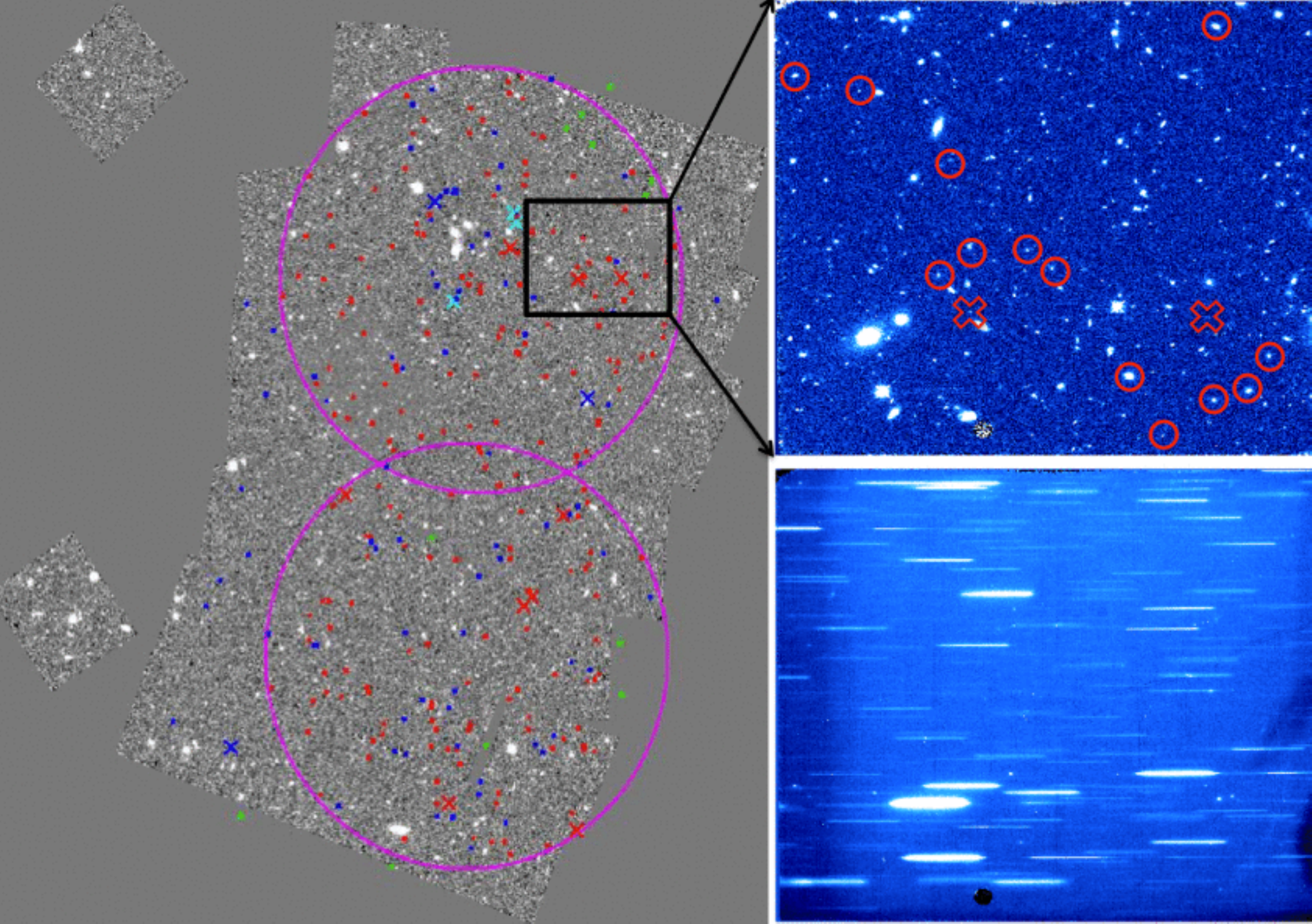}
\caption{Direct image mosaic map (comprising 38 pointings) of the GOODS-S (RA: 3$^h$ 32$^m$ 30$^s$, Dec: -27$^d$ 48$^m$ 54$^s$) region. The size of each pointing can be seen from the two outlying fields on the left of the image. The image shows the WFC3-ROLES galaxies and ESO Public spectroscopy galaxies which have a spectrum covering the \ha~wavelength range (red and blue points respectively) and the galaxies that do not have a spectrum (red and blue crosses respectively) covering the wavelength range where \ha~is expected. Galaxies not reduced by the spectroscopic reduction pipeline are indicated by cyan crosses. Galaxies which fall in the gaps between pointings (green circles) are excluded from the WFC3-OII sample. The ROLES footprint is indicated by two magenta circles of roughly 8 arcminute diameter. The insets show a direct image (red circles show WFC3-ROLES galaxies with spectra and red crosses show those without spectra) and the corresponding 2D slitless spectrum of a single pointing.} 
\label{mosaic_map} 
\end{figure*}

\noindent The GOODS-S region is an extragalactic field well studied by many surveys, one of which is \tdhst.\footnote{\url{http://3dhst.research.yale.edu/}}
The \tdhst~survey obtained low resolution NIR slitless spectroscopic data from observations taken with the WFC3/G141 grism from \textit{HST} together with WFC3/F140W direct images.  The GOODS-S field comprises 38 pointings, as shown by the direct mosaic map in Fig.~\ref{mosaic_map}. Fig.~\ref{mosaic_map} shows the layout of the ROLES field with the WFC3-OII galaxies overlaid, together with a zoom-in of a single pointing and its corresponding slitless 2D spectroscopic image. We  target the H$\alpha$ emission line in our \oii-selected sample of  WFC3-OII galaxies. There are 12 galaxies that fall in gaps between pointings, which reduces our sample to 299 WFC3-OII galaxies.

 The WFC3/G141 spectra have a wavelength coverage from approximately 11000\AA~to 16500$\textrm{\AA}$ at a spatial resolution of $\sim$0.13 arcsec. \citep{2012ApJS..200...13B}. The mean dispersion of the primary spectral order of the G141 grism is 46.5$\textrm{\AA}$/pixel and the size of the resolution element is $\sim$100\AA~(\textit{R}$\sim$120 at 13000\AA). These grism specifications enable the detection of the \ha~emission line in the redshift range 0.7 $<$ \z~$<$ 1.5. 

In this paper, we use the SED-fit SFRs, dust extinction estimates and stellar masses from ROLES. They obtained these quantities by fitting deep multiwavelength photometry at each galaxy's spectroscopic redshift to a grid of stellar population synthesis (SPS) models (using PEGASE.2; \citealt{1997A&A...326..950F}) as described in \cite{2004Natur.430..181G}.

\subsection{Data Reduction and Spectral Extraction}
We reduce the \textit{HST} spectroscopic data using the spectral extraction software package, \texttt{aXe} \citep{2009PASP..121...59K}. Details of the procedure are given in \cite{2012ApJS..200...13B}. Briefly, the task \texttt{Multidrizzle} \citep{2006hstc.conf..423K} combines the multiple direct and corresponding slitless spectra from each pointing to obtain a deep exposure direct image and corresponding 2D slitless spectrum. The object positions are obtained by running the object detection algorithm, \texttt{SExtractor}, \citep{1996A&AS..117..393B}  on each combined direct image. A global background subtraction is done whereby the 2D master sky image is scaled to the background of the 2D slitless spectra and then subtracted. In slitless spectroscopy, contamination from overlapping spectra can be problematic. In \texttt{aXe}, the quantitative contamination method is used to determine the level of contamination in each spectrum. An estimate of the contaminating flux from all other sources is determined for each spectral bin using emission models. The dispersed contribution of every object to the slitless spectrum is modeled and, using the model information, the contaminating flux for each pixel is recorded and processed through the extraction process. This results in a contaminating flux spectrum for each extracted spectrum. Out of our selected sample of 299 galaxies a total of 281 spectra are successfully extracted. There are 15 spectra that do not cover the wavelength range where \ha~is expected, due to their location near the edges of the detector, and a further three spectra that the \texttt{aXe} pipeline fails to extract. Fig~\ref{lie} shows an example of a 2D and 1D spectrum extracted by the \texttt{aXe} pipeline. 
\\

\section[]{Spectral Analysis}
\label{sec:spec}
We use an automated algorithm to verify the presence and measure the luminosity of the \ha~emission line for each spectrum. For this dataset, we have the ROLES \oii-based spectroscopic redshifts, which means that we know where to search for the expected wavelength of H$\alpha$. 

However, we do not always expect to detect an emission line. This is because the galaxies we are targeting have very low masses and therefore low SFRs ($\sim$0.1 \Msunyr). We can estimate the approximate limiting SFR of the H$\alpha$ data by using the standard relation from \citet{1998ARA&A..36..189K} (converted to our assumed IMF) and setting this equal to the typical flux limit (3$\sigma$) of the HST spectra ($\sim$3$\times$10$^{-17}$ erg/cm$^2$/s). This equates to a SFR limit of $\sim$1 \Msunyr. This is a factor of three higher than the nominal ROLES' limit for \oii~which means that we do not expect to detect the lowest SFR galaxies. In addition, these galaxies are so faint that we do not expect to detect continuum for most of them.
Thus, the \textit{HST} spectra for our dataset are unusual in the sense that they are mostly noise (undetected continuum), but with a fraction exhibiting a significant emission line. Furthermore, these data are unusual in that they are very low resolution spectra (100\AA), meaning that the line is unresolved. In fact, the resolution element is so wide that the \nii$_{\lambda6548}$ and \nii$_{\lambda6583}$ lines are blended with the \ha~line. To correct for the flux contribution from \nii, we assume a typical ratio of \nii/(\nii$+$H$\alpha$) $\approx$ 0.25 \citep{2013MNRAS.428.1128S}.

\subsection{Automated Line Detection Algorithm}
\begin{figure}
\noindent \begin{centering}
\scalebox{0.45}{\includegraphics{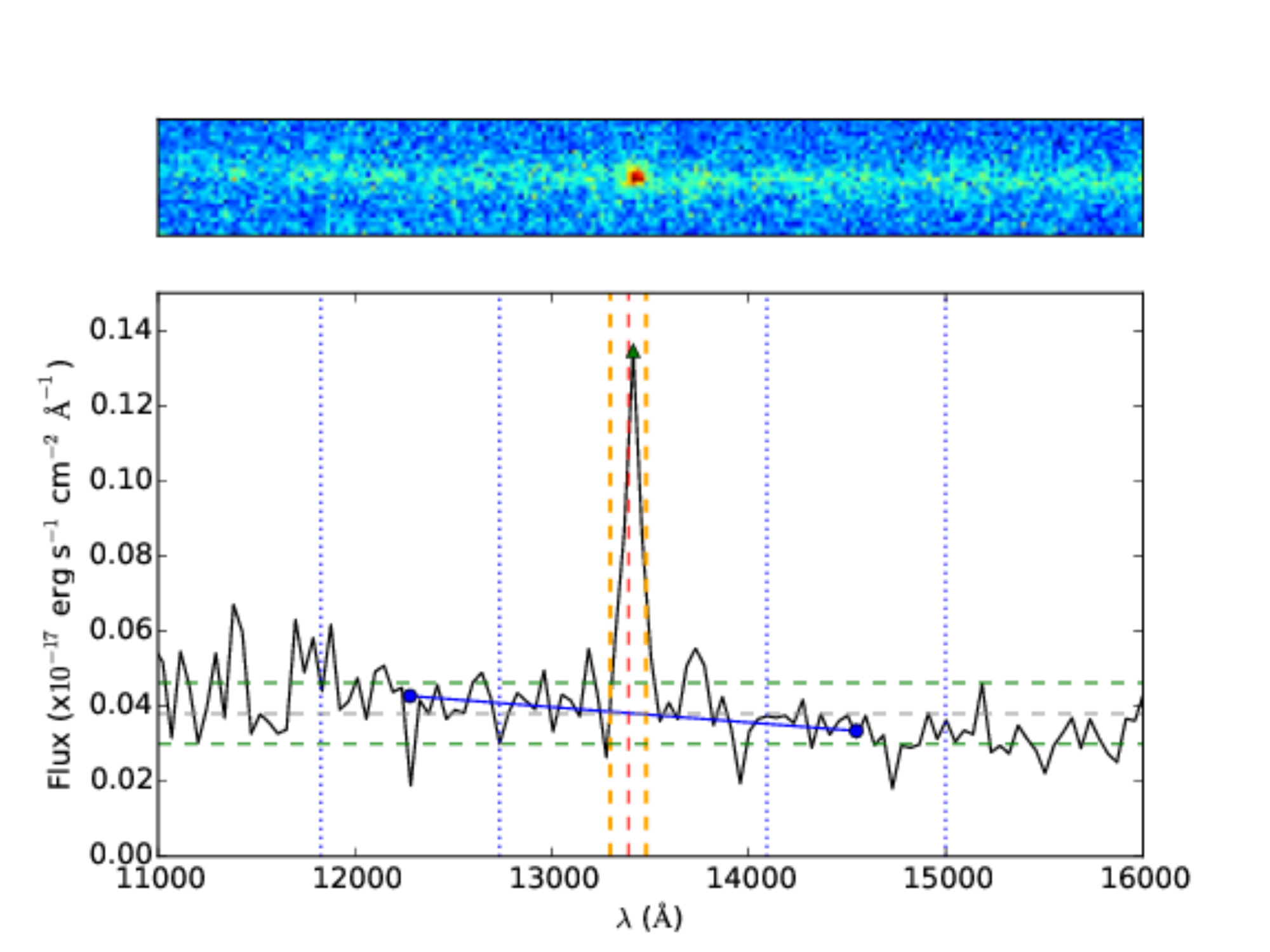}}
\par\end{centering}
\caption{An example of a galaxy 2D (upper panel) and 1D spectrum extracted by the aXe pipeline. Bottom panel: The expected position of the emission line (dashed red line) around which the 4 pixels ($\sim$200$\textrm{\AA}$) window (dashed orange lines) was centered. The emission line peak  (green triangle) was found within this window. The sidebands are indicated by the blue dotted lines. The blue solid line is the fit to the continuum obtained from the mean flux in the sidebands (blue circles). The green dashed lines indicate the 1$\sigma$ noise level for a flat continuum (grey dashed line). The smaller peak around 13700 $\textrm{\AA}$ is consistent with redshifted SII at $z=1.04$. See text for discussion.}
\label{lie}
\end{figure}

We briefly outline the steps our automated algorithm uses to detect and measure \ha~flux. The aspects of the 1D spectra discussed below are illustrated in the lower panel of Fig.~\ref{lie}

\subsubsection{Line Verification}
\label{sect:lineveri}
To account for uncertainty in the precise location of the emission line due to wavelength calibration of the G141 grism or the calibration used by ROLES to determine the galaxy redshifts, the algorithm begins by placing a window $\sim$200$\textrm{\AA}$ (4 pixels) wide centered on the expected line position. The final line position was identified as the pixel containing the highest flux within the window.\footnote{The (internal) precision of the ROLES \oii~redshifts is 100 \kms~observed frame \cite{2010bMNRAS.405.2419G}, so in principle a one WFC3 pixel window would be sufficient. However, since we use a 10 pixel window to measure the integrated flux (described below) in practice it makes negligible difference to the final flux whether we use a one or four pixel centring box.}

\subsubsection{Continuum Estimate}
\label{sect:cont}
The flux at the position of the line contains flux from both the line and the continuum. The latter must be estimated and subtracted from the total flux in order to determine the true line flux. This is done by calculating the continuum in two 1D sidebands on either side of the line. A width of 20 pixels, ranging from ($x_{peak}$ -15) pixels to ($x_{peak}$ - 35) pixels and ($x_{peak}$+15) pixels to ($x_{peak}$ + 35) pixels, is chosen as a reasonable estimate for the sidebands. These values were chosen by visually inspecting all spectra which showed obvious emission lines, to decide firstly, how far from the line to position the sideband so that the line flux was not included when calculating the continuum and secondly, how wide to make the sidebands to ensure that they contained enough continuum flux such that a linear fit to the continuum was valid.

\subsubsection{Line Luminosity Measurement}
\label{sec:linelum}

The final step in the algorithm is to measure the line luminosity and its uncertainty. The line has some width, so to calculate the total flux of the line, we have to integrate over a finite region. Choosing an integration region was a trade off between including too much noise by making the region too wide and not including all the emission line flux by making the region too narrow. To determine how wide a region to integrate over, spectra were visually inspected and a width of 480$\textrm{\AA}$ (10 pixels), centred on the peak of the emission line, was chosen as a reasonable estimate. The line luminosity was then calculated using
\begin{equation}
L(H\alpha)=4\pi F(H\alpha)(D_{L}^{2})
\end{equation}
\newline
where $F(H\alpha)$ is the integrated flux of the emission line after continuum subtraction and D$_{L}$ is the luminosity distance (e.g. \citealt{1999astro.ph..5116H}).

The uncertainty on the flux, $\sigma$ is calculated as the standard deviation integrated over a region the same width as the integration region of the emission line (10 pixels or 480\AA). The noise just takes into account noise from the continuum and not Poisson noise from the emission line itself. For bright lines, the noise will be mis-estimated, but here we care primarily about the detection of faint lines where the Poisson noise from the line itself will be low.

\subsubsection{S/N of detections}

In order to determine the significance of detections, we adopt a slightly more conservative approach than for the flux measurements. We define the signal-to-noise relative to the peak pixel in the detection. The S/N ratio is then calculated as 

\begin{equation}
$\sn$ = \frac{F_{\rm peak}(H\alpha)}{\sigma_{\rm per\,pixel}}
\label{sneqn}
\end{equation}
where $F_{\rm peak}(H\alpha)$ is the flux in the peak pixel of the emission line after continuum subtraction and $\sigma_{\rm per pixel}$ is the noise per pixel of the spectrum, calculated from the side bands as above. 

As described above, our estimate of \sn~is mostly used as an approximation to rank order detections, rather than a detailed estimate of the noise (as calculated from the continuum noise above, for example). As with ROLES, we found the best method to assess significance of detections was a combination of visual inspection and reproducibility of detections in repeated observations.

\begin{figure}
\noindent \begin{centering}
\includegraphics[width=0.5\textwidth]{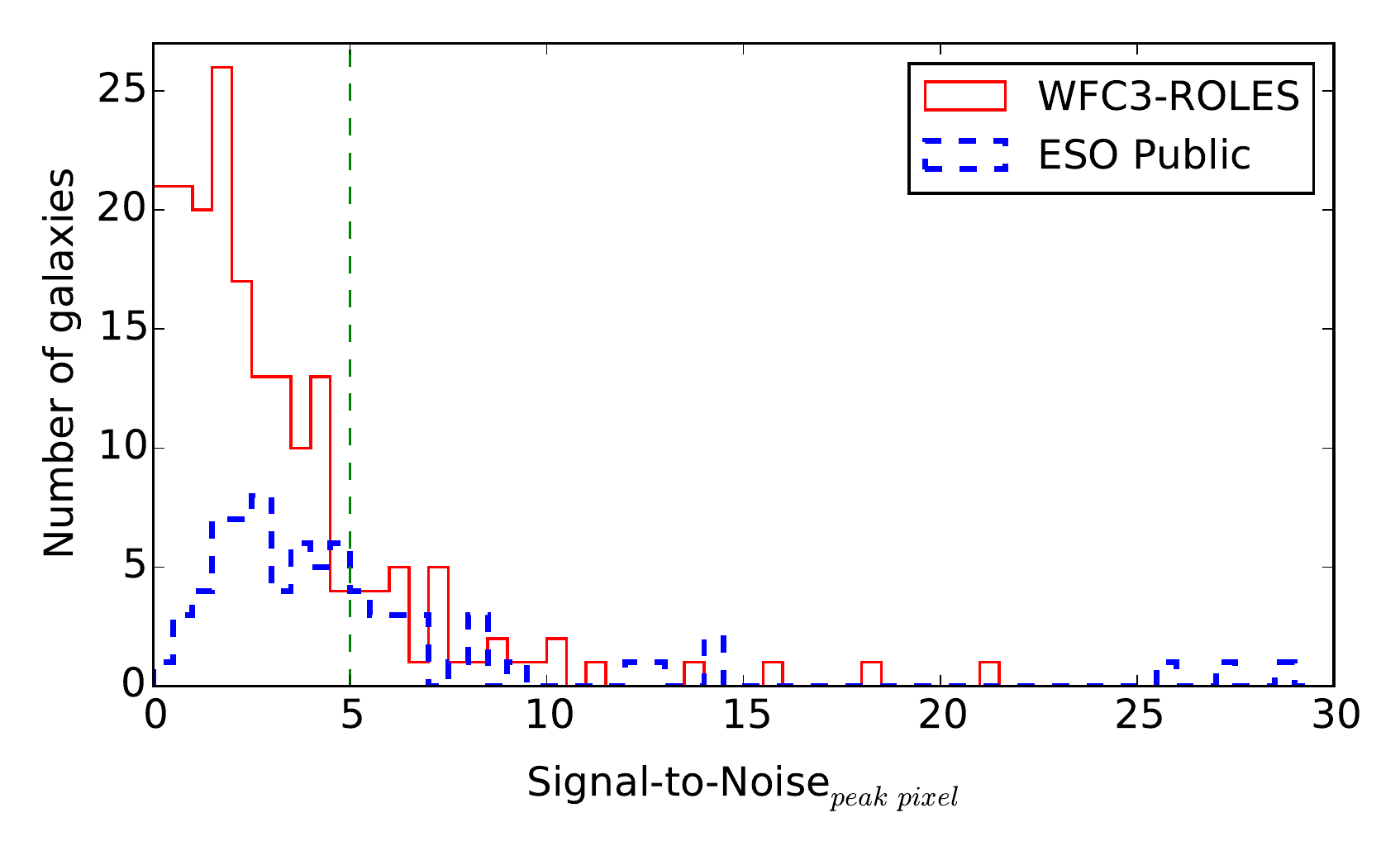}
\par\end{centering}
\caption[Caption]{Distribution of the significance of the emission lines for all 281 unique galaxies
in the sample separated into WFC3-ROLES (red) and ESO Public Spectroscopy galaxies (blue). This shows the relative number of detections and limits expected for any chosen \sn~threshold (the green dashed line shows our chosen detection threshold of \sn$\geq$5). Objects with a \sn~greater than the threshold are considered detections and those less than the threshold are limits. }
\label{hist}
\end{figure}

As mentioned before, we do not always expect to detect an emission line. For this reason, a detection threshold has to be defined for our sample. Choosing a detection threshold is a trade off between purity and completeness. A very high threshold means that one has a pure sample which is incomplete (i.e missing real detections). On the other hand, as one moves to lower thresholds, the sample becomes more complete but the probability of including spurious detections increases. 
A S/N $\geq$ 5 is chosen as the detection threshold from a combination of visual inspection and testing the reproducibility of automated detections (detailed in Appendix A). The distribution of \sn~of the 281 WFC3-OII galaxies is shown in Figure \ref{hist}. Using this threshold we obtain 56 detections.

\section{Results and Discussion}

\subsection{Dust Extinction Correction}

In order to convert from a measured \ha~luminosity into a SFR it is necessary to have an estimate of the dust extinction. Unfortunately, our spectra do not cover the wavelength range of the H$\beta$ emission line,  which means that the dust extinction cannot be measured using the Balmer decrement (H$\alpha$/H$\beta$) as is commonly done (e.g., \citealt{2013ApJ...777L...8K}). Instead we use SED-fits to estimate the dust extinction, $A_V$, and compare our measurements to other studies where the dust extinction has been measured using the Balmer decrement. Since dust is created in stars, one might expect dust extinction to depend on the number of stars (stellar mass), and/or the rate at which they are formed (SFR),  in a galaxy.  Previous studies (using \ha/H$\beta$ to measure dust) have shown that dust extinction correlates with stellar mass (e.g. G10, \citealt{2010MNRAS.409..421G}, \citealt{2012MNRAS.420.1926S}, \citealt{2013ApJ...777L...8K}, \citealt{2013AJ....145...47M}) and H$\alpha$ luminosity (or SFR) (e.g. \citealt{2001AJ....122..288H}, \citealt{2003ApJ...591..827P}, \citealt{2005ApJ...619L..51B}, \citealt{2006ApJ...643..173S}, \citealt{2008ApJ...680..939C}).

\begin{figure*}
\noindent \begin{centering}
\includegraphics[width=1.0\textwidth]{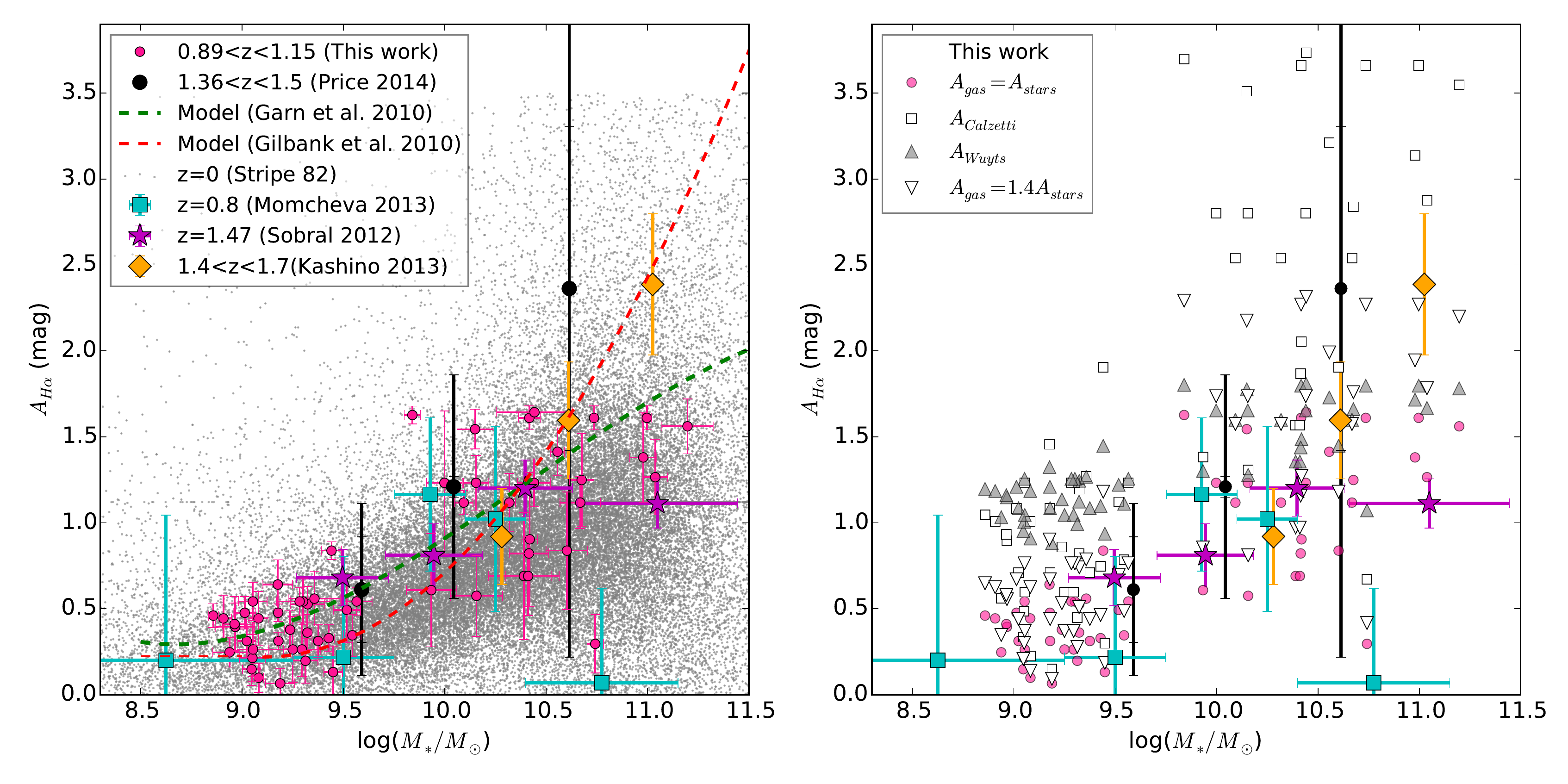}
\par\end{centering}
\caption{The relation between dust extinction (A$_{H\alpha}$) and stellar mass. All external comparison samples estimate dust using the Balmer decrement. Our SED-fitted dust estimates (pink points) for the low mass galaxies agree well with the Momcheva (2013) data at a similar redshift and also with local relations derived from SDSS data. Small, grey points show (Balmer decrement based) measurements from SDSS. The right panel repeats the z$\sim$1 observations from the left panel (coloured points) and compares with four different scalings of dust extinction for our data (grey points, error bars omitted for clarity). See text for discussion. 
\label{extinc_mass}
}
\end{figure*}

We calculate extinction assuming a \citet{2000ApJ...533..682C} law for the continuum, $A_V$ ($A_{\rm stars}$), and SMC for the nebular emission, H$\alpha$ ($A_{\rm gas}$) \citep[e.g.,][]{2014ApJ...795..165S}.  While Calzetti et al.\ advocate a ratio between $A_{\rm stars}$ and $A_{\rm gas}$ of 0.44 to allow for the differential attenuation between the two regions, i.e.,

\begin{equation}
A_{H\alpha} = \frac{k_{H\alpha}}{k_V} \frac{A_V}{0.44} = 1.8673 A_V,
\label{dust}
\end{equation}
(where  $k_{H\alpha}$=3.3258 and $k_{V}$ =4.04789), other works argue that the factor of 0.44 appearing in the denominator of Eq 3 may depend on the properties of the galaxies and growing evidence at z$\sim$1 suggests that the additional extinction suffered by the gas relative to the stars should be lower than the Calzetti value. For example, \citet{2014ApJ...788...86P,Reddy2015:apj259,Wuyts2013:apj135} with some values ranging as low as no additional extinction \citep[e.g.,][or A$_{\rm gas} = A_{\rm stars}$]{2013ApJ...777L...8K}.

\begin{equation}
A_{H\alpha} = 0.822 A_V.
\label{eqn:usedust}
\end{equation}
For simplicity, we begin by assuming equal extinction between stars and gas as a lower limit, but explore some of the alternative suggested parameterisations as a function of stellar mass. In Fig.~\ref{extinc_mass} we plot our SED-fitted dust estimates using Eq.~\ref{eqn:usedust} as a function of stellar mass. For comparison, the median dust extinction and stellar mass values of \cite{2013AJ....145...47M},  \cite{2012MNRAS.420.1926S} and  \cite{2013ApJ...777L...8K} are plotted together with the local relations derived by \cite{2010MNRAS.409..421G} and G10 at \z$\sim0.1$ given by the following respectively:

\begin{equation}
A_{H\alpha} = 0.91 + 0.77 \hskip 0.07cm X + 0.11 \hskip 0.07cm X^{2} - 0.09 \hskip 0.07cm X^{3}~{\rm (Garn~\&~Best~2010)}
\label{garndust}
\end{equation}
where \textit{X}=log(M$_{*}$/10$^{10}$M$_{\odot}$), and

\begin{equation}
A_{H\alpha}=a + b \hskip 0.07cm log(M_{*}/M_\odot) + c \hskip 0.07cm [log(M_{*}/M_\odot)]^2 \hskip 0.5cm {\rm (G10)}
\label{dust2}
\end{equation}
where $a$=51.201, $b$=-11.199, $c$=0.615 and is set to a constant value for log(M$_{*}$/M$_{\odot}$) $\leq$ 9.0.

The \cite{2013AJ....145...47M} sample at \z$\sim$0.8 is the closest to our redshift range. For low mass galaxies (log$(M_{*}/M_{\odot})$ $<$ 10) our dust estimates agree well  with the median Balmer decrement dust measurements of \cite{2013AJ....145...47M} within the uncertainties. Furthermore, our dust estimates for low masses fall on the local relations of \cite{2010MNRAS.409..421G} and G10, both derived using SDSS data. At the highest masses however, our measurements begin to fall below the G10 relation locally and also the highest mass data point at z$\sim$1.5  \cite{2013ApJ...777L...8K}. The highest mass point of \citet{2013AJ....145...47M} is offset even further below the local relation. They mention that this data point may have possible contamination by their stacked sample due to unidentified AGN at this mass range. However our data agree well with the Sobral et al., measurements at z$\sim$1.45 and are consistent with the Garn (2010) local relation.

In the right panel of the Figure, we explore some of the alternate parametrisations of \aha~as a function of stellar mass. In addition to the \citet{2000ApJ...533..682C} relation and our equal extinction in stars and gas relation (essentially $0.44A_{H\alpha,Calzetti}$); we use the relation from \citet{Wuyts2013:apj135}, $A_{\rm gas} = 1.9 A_{\rm stars} - 0.15 A_{\rm stars}^2$; and $A_{\rm gas} = 1.4 A_{\rm gas}$ which is an approximation to the relation found by \citet{Reddy2015:apj259}. The latter relation seems to bring our data points closest to the observed Balmer decrement measurements found in other works, but $A_{gas} = A_{stars}$ is also in reasonable agreement, if somewhat lower than \citet{2013ApJ...777L...8K} at the highest masses. Using \citet{Wuyts2013:apj135} relation, our lowest mass galaxies significantly exceed the extinction predicted by other works \cite{2013AJ....145...47M,2012MNRAS.420.1926S}; and the Calzetti correction significantly overpredicts the extinction in the highest mass galaxies, as found by other works. For simplicity, we choose to use $A_{\rm gas} = A_{\rm stars}$ (Eqn.~\ref{eqn:usedust}) throughout, but will discuss the differences these other reasonable choices, particularly $A_{\rm gas} = 1.4 A_{\rm stars}$ would make to our results.

\subsection{SFR Indicators}

In this section we examine all the SFR estimators available for use with our data. These comprise three indicators (\ha, \oii, and SED-fitting), each with various possible assumptions and corrections. Each of these has its own advantages and disadvantages, including different sensitivities to factors such as dust, metallicity, as well as the timescales to which it is sensitive. A more thorough discussion can be found in G10. 

As we have shown in the previous section, using $A_{H\alpha}$ estimated from the SED-fitted continuum extinction, scaled as described in \ref{eqn:usedust} gives reasonable agreement with the Balmer decrement based $A_{H\alpha}$ values from other studies at similar redshifts. Thus we use this as our fiducial scaling and test the calibration of other available indicators against this.

\subsubsection{\ha~\rm{SFR}}
We calculate the \ha~SFR using,
\begin{equation}
\rm{\frac{SFR}{[M_{\odot} \hskip 0.07cm yr^{-1}]}=\frac{10^{0.4A_{H\alpha}}}{1.5}\frac{L(H\alpha)}{1.27\times10^{41} [erg \hskip 0.07cm s^{-1}]}}
\label{halpha1mag}
\end{equation}
\noindent where $A_{H\alpha}$ is calculated using Eq. \ref{eqn:usedust} and L(H$\alpha$) is the measured H$\alpha$ line luminosity (see \S \ref{sec:linelum}). This is the \citet{1998ARA&A..36..189K}, K98, relation divided by a factor of 1.5 to convert from a Salpeter IMF to a \cite{2001MNRAS.322..231K} IMF.

\subsubsection{\rm{Nominal \rm{\oii}~SFR} (\rm{\oii}K98)}
The SFR measured from \oii~luminosity was calibrated by  K98 by scaling between the \oii~and H$\alpha$ luminosity as:
\begin{equation}
\rm{\frac{SFR}{[M_{\odot}yr^{-1}]}=\frac{L(\oii~)}{2.53\times10^{40} [erg s^{-1}]}}
\label{nomoiieqn}
\end{equation}
The original scaling was determined by K98 from the \ha~SFR and so the normalisation constant assumes nominal values for the extinction and ratio of \ha/\oii~luminosity, etc. Our intention here with the \oii K98 method is to just use the nominal scaling from \oii~luminosity, as would be used in the absence of other measurments such as extinction, metallicity, etc., to see how well this compares with the other estimators. Due to the relative ease of obtaining \oii~SFRs versus \ha~at these redshifts, \oii~is still an important tracer of SFR (see G10 for further discussion).\\

\subsubsection{\rm {Empirically corrected} \rm{\oii}~\rm{SFR} (\rm{\oii}G10)}
\label{sect:SFRG10}
The empirically corrected \oii~SFR (referred to as \oii G10) is calculated using the G10 mass-dependent empirical correction,
\begin{equation}
\rm{\frac{SFR_{emp,corr}}{[M_{\odot} \hskip 0.07cm yr^{-1}]}}=\frac{\rm{L({[}O II{]})}}{2.53\times 10^{40} }\frac{1}{(a\tanh[(x-b)/c]+d)}
\label{massdep}
\end{equation}
\noindent where $x$=$\log(M_{*}/M_{\odot})$,  $a$=-1.424,  $b$=9.827, $c$=0.572,
$d$=1.700  and L({[}O II{]) is the \oii~emission line luminosity.

As mentioned, the nominal luminosity scaling of \oii K98 method is sensitive to various factors such as dust extinction, metallicity, ionisation parameter, etc. G10 attempted to correct for these by parameterising the combined dependence of all these factors empirically as a function of stellar mass.

\subsubsection{SED-fit SFRs}
SED-fit SFRs are calculated from the same procedure as our stellar masses. The method used is that detailed in \citet{2004Natur.430..181G}, fitting the extensive multiwavelength photometry (see G10) at the spectroscopic redshift to a grid of PEGASE.2 \citep{1997A&A...326..950F} stellar population models.

\subsection{\ha~and \oii~SFR Comparison}
\label{sec:sfrs}
\begin{figure*}
\centering
\includegraphics[width=18cm]{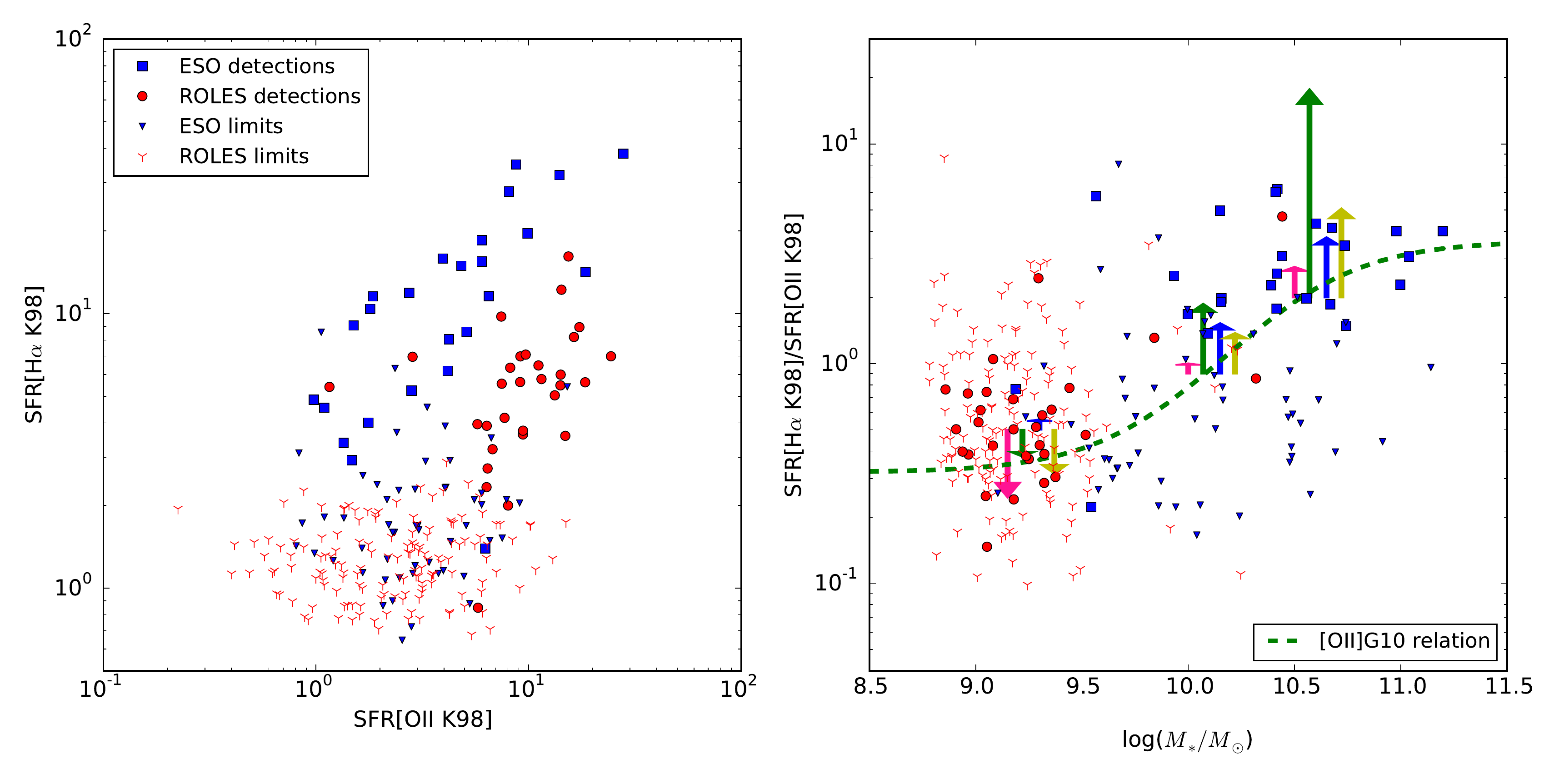}
\includegraphics[width=18cm]{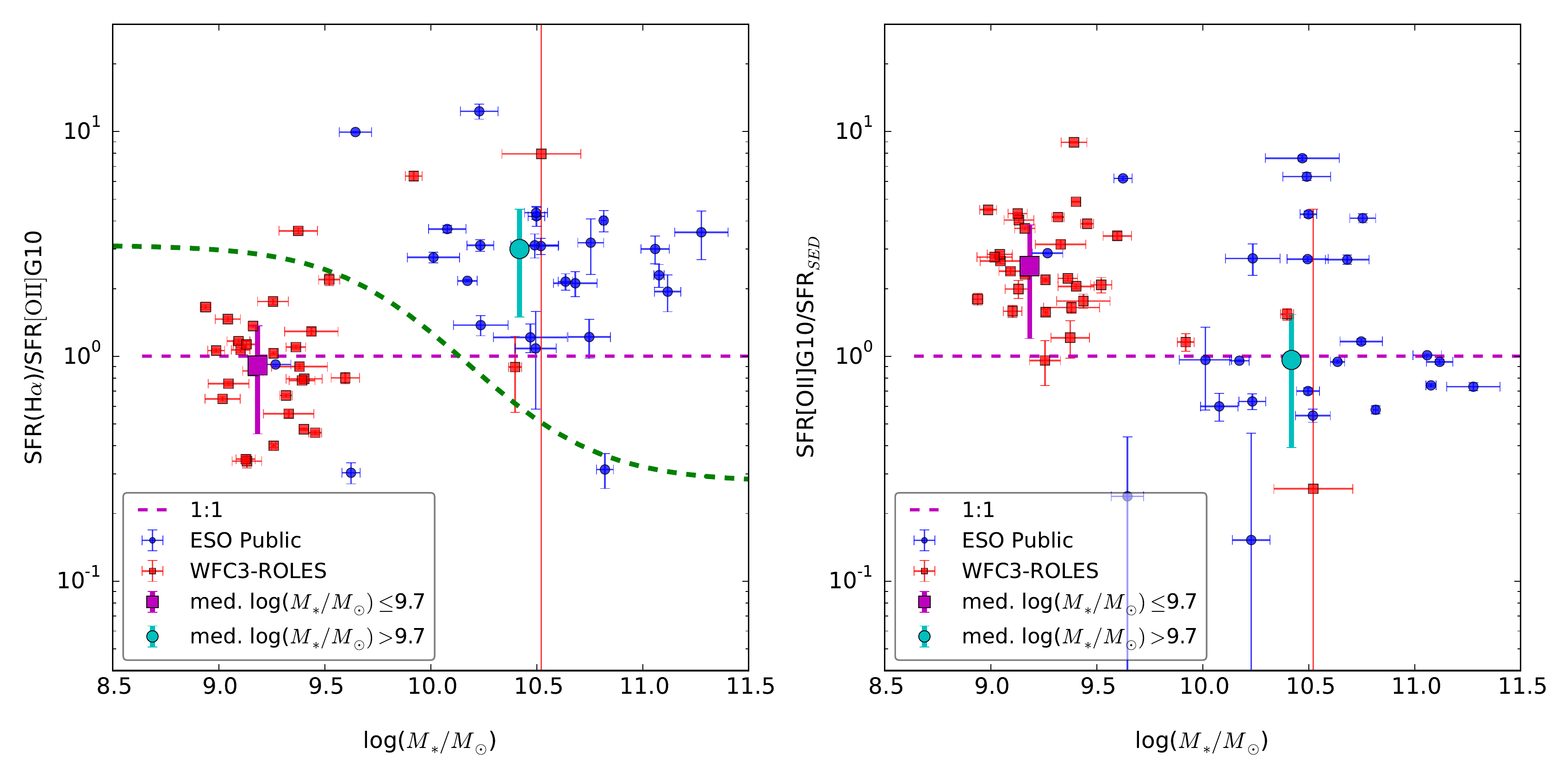}
\caption{Upper left panel: the comparison between SFRs directly scaled from luminosity (K98) for \ha~and \oii. Larger symbols indicate \ha~detections and smaller symbols denote upper limits in \ha. Upper right panel: the ratio of the K98 SFRs as a function of stellar mass, symbols as previous panel. The dashed relation indicates the empirical mass-dependent correction for \oii~SFR (G10). Arrows indicate the change in SFR by adopting the different dust correction schemes (from left to right: $A_{\rm gas} = A_{\rm stars}$, $A_{\rm Calzetti}$, $A_{\rm Wuyts}$, $A_{\rm gas} = 1.4 A_{\rm stars}$) for three representative mass bins. See text for discussion. Lower left panel: The ratio of the (extinction-corrected) H$\alpha$ and \oii G10 SFR relations as a function of stellar mass. The G10 empirical correction fixes most of the mass-dependent trend which would be present if only a constant conversion from \oii~luminosity to SFR is used (dashed green line). Lower right panel: The \oii G10 SFR and SED-fitted SFR ratio as a function of stellar mass shows a small residual trend with mass. See text for discussion.}
\label{sfratio}
\end{figure*}

In Fig.~\ref{sfratio}, we begin by comparing the nominal (K98) scalings between \ha~and\oii. These are essentially direct scalings from line luminosities and so are closest to the raw data. The top left panel shows a correlation between the two line luminosities albeit with significant scatter. In addition, for the detections (larger points) there seems to be a systematic offset between the ROLES (lower mass) and ESO (higher mass) measurements. In the upper right panel, the ratio of the line luminosities is shown as a function of stellar mass. The mass dependent offset is now clearer, and as can be seen from the dashed line, the empirical correction to the \oii~SFR (G10) corrects for the bulk of the offset prior to any additional corrections (such as mass-dependent extinction) in the \ha~data, although the scatter is still broad. To illustrate the effect of the different dust corrections explored in the previous section, the arrows indicate the change in \ha~SFR for typical nominal \ha~SFRs at three representative masses, relative to the K98 $A_{\rm H\alpha}$=1.0 mag correction. It can be seen that the dust-corrected \ha~SFRs are reduced at the lowest masses, and increase by differing amounts at higher masses, with the Calzetti correction leading to the highest increase, and our adopted $A_{\rm gas} = A_{\rm stars}$ giving the smallest increase. In the highest mass bins, the agreement between our adopted measurement and the two other alternatives is much closer than their agreement with the Calzetti relation.

In the lower left panel of Fig.~\ref{sfratio} we plot the ratio of the (extinction-corrected) \ha~and the empirically corrected \oii G10 SFRs as a function of stellar mass. The dashed line shows where the data would fall if this correction is not applied and the nominal (constant) conversion from \oii~luminosity to SFR (K98) is used instead. Clearly this would leave a residual mass trend between the high and low mass galaxies, as is also seen from the upper right panel: most of the discrepancy between \ha~luminosity-inferred SFR and \oii~luminosity-inferred SFR is reconciled by this mass-dependent correction. The motivation for this correction is given in G10 and the likely applicability at z$\sim$1 is discussed in \citet{2010bMNRAS.405.2419G}. Briefly, 
the conversion from \oii~luminosity to SFR depends on gas phase metallicity, dust extinction, and ionisation parameter (e.g., \citealt{2004AJ....127.2002K}). Several workers have invoked empirical corrections to \oii~SFRs based on broad band luminosities or stellar masses. In as much as the mass dependent scaling of these various contributions remain approximately constant (or even evolve but in opposing directions so as to cancel out), the empirical relation derived at z$\sim$0.1 might also be expected to work at z$\sim$1.0. A full exploration of the applicability of this correction is difficult, given the limited data available (individual galaxies spanning a range of stellar masses with \oii, \ha~luminosity measurements and reliable individual dust measurements), but we can estimate how much the G10 relation might evolve by z$\sim$1 by looking at the measured evolution in the mass--metallicity relation (e.g., \citealt{Savaglio2005:apj260}) and the mass--extinction relation (e.g., Fig.~\ref{extinc_mass}).\footnote{Remember: these components are not fitted individually in the G10 derivation, only the overall contribution of all terms including ionisation parameter, for which we do not have measurements for these galaxies.} From \citet{Savaglio2005:apj260}, at a fixed metallicity, the corresponding stellar mass changes by  \logmass$\lsim$0.4 between the local Universe and their high redshift bin (z$\sim$0.7). The bulk of the high and low mass galaxies in this sample lie on the flat parts of the tanh correction curve, and so choosing a correction from \logmass$\sim$0.4 away makes little difference to the correction factor. Or, alternatively, metallicity decreases by $\sim$0.1--0.2 dex at fixed stellar mass up to this redshift. The decrease in metallicity means a higher \oii/\ha luminosity at a given SFR (e.g., \citealt{2004AJ....127.2002K}), implying a given \oii~luminosity would overestimate the SFR using a fixed luminosity conversion. This may be partly cancelled by an increase in dust extinction relative to z$\sim$0 (Fig.~\ref{extinc_mass}). Thus it is not unreasonable that the local empirical correction could also apply at z$\sim$1. Its original motivation was to correct for biases in measurements of SFR and SFR density at z$\sim$1 as a function of stellar mass using \oii~as the only SFR indicator \citep{2005ApJ...619L.135J}, and so part of the motivation was to use a mass-dependent correction since this quantity was always available as part of the studies. Whether it is the best or observationally cheapest proxy for such studies is a separate question. \citet{Reddy2015:apj259} favour SSFR based scalings in order to infer dust corrections at z$\sim$1.5, and so similar scalings will be important in the near future in order to maximally extract information from such studies where individual estimates of all relevant quantities may not be available individually for each galaxy.

Most of the WFC3-ROLES data (almost exclusively low mass galaxies, by design) lie below the K98 relation whereas most of the ESO Public data (higher mass galaxies) lie above the K98 relation. The median ratio of the lower mass data (mostly WFC3-ROLES data) is 0.29$\pm$0.15 (0.80$\pm$0.36) without (with) this empirical correction and the median for the higher mass data (mostly ESO Public data) is 3.59$\pm$2.78 (1.94$\pm$1.01) without (and with correction, respectively). There is more scatter in the high mass data than the low mass data which means that the data is less well constrained at high masses.  Most interesting is that the ratio as a function of mass follows the mass dependent correction of G10 derived at low redshift (Eq. \ref{massdep}, purple line) within the uncertainties. This shows us that it is necessary to apply the mass-dependent \oii G10 correction to \oii~SFRs rather than simply assuming a constant \oii~luminosity to SFR conversion.  Furthermore, this confirms that the \oii G10 relation at z$\sim$1 is consistent with the normalisation found at z$\sim$0.1. We cannot however, rule out moderate evolution in this correction given the broad uncertainties of our z$\sim$1 \ha~measurements.

\subsection{Comparison with SED-fit SFRs}
We now compare the ratio of the SED-fitted SFRs and \oii SFRs as a function of stellar mass in the right panel of Fig.~\ref{sfratio}. (The ratio between \ha~and SED fit SFRs can similarly be inferred from the combination of the two, given the agreement between \oii G10 and \ha.) The ratio of the SFRs for the high mass bin is consistent with unity (1.0$\pm$0.9), however, for the low mass bin the ratio is 0.4$\pm$0.2, suggesting a slight residual trend for SED-fit SFRs relative to \oii~G10 or \ha. This is in contrast to \cite{2010bMNRAS.405.2419G} who found that the SED-fitted SFRs are approximately 1.7 times higher than \oii G10 for ROLES. This could be due to sample selection in the current comparison. Since we only compare objects with significant \ha~detections, if the \oii(or \ha)--SFR relation has moderate scatter, an Eddington-like bias could lead us to pre-select only the high \ha(/\oii) outliers in \ha~vs SED in this comparison.  

Other studies have used the \oii G10 empirical correction. \cite{2012ApJ...746..124M} showed how \oii G10 correction compares to SFRs measured at z$\sim$1 using SFR fits from the UV/optical SEDs in the AEGIS survey. They found that the SED-fitted SFRs and the \oii G10 empirically calibrated SFRs agree with 0.27 dex (1.86) scatter and have a mean offset of  -0.06 dex (0.87). \cite{2013MNRAS.431.1090M}  used the \oii G10 correction to obtain SFRs for a sample of galaxy groups from the Group Environment Evolution Collaboration 2  (GEEC2; Balogh 2011). They compared the SFRs derived from this method to SFRs calculated from using FUV SED-fits combined with  24$\mu$m, which should trace the total (obscured and unobscured)  star formation in the galaxy. They noticed a systematic normalization offset between the two indicators. They found that the \oii~SFR estimated using \oii G10 is underestimated by a factor of 3.1, much larger than the offset found here and by the other cited works.

\subsubsection{Possible systematic uncertainties}
The two main sources of uncertainty in our \ha~SFRs are the \nii~correction to the blended (\ha+\nii) flux and the correction for dust extinction. We used an average \nii/(\ha+\nii) correction of 0.25 which corresponds to an \nii/\ha~of 0.33 and is close to the values derived by \citet{2012MNRAS.420.1926S} and \citet{2013MNRAS.428.1128S} at these redshifts, and is the same as the conventionally used 33$\%$ in \nii/\ha. Some studies have shown that the \nii/\ha~ratio increases with stellar mass (e.g. \citealt{2006ApJ...644..813E}). Furthermore the ratio, at any particular stellar mass, evolves with redshift (e.g. \citealt{2006ApJ...644..813E}).  If the ratio has been overestimated for the low mass galaxies and underestimated for the high mass galaxies, it could result in the slight residual trend seen in our data in Fig.~\ref{sfratio}, left panel. Typically used \nii/\ha~ratios are in the range 0.3-0.5 (\citealt{1983AJ.....88.1094K}, \citealt{1992ApJ...388..310K}). Applying the extremes of  the ratio to the data, bring the \ha~and [OII] SFR ratios close to unity as possible but there still is a residual trend. Our result of a mass-dependent difference between \ha~and \oii~is therefore robust to any reasonable choice of \nii/\ha~correction. \\

In order to investigate the effects that our choice of dust extinction estimate has on the differing ratios of SFRs for high and low mass galaxies we look at a range of possible extinction estimates by considering the alternative dust extinction laws discussed in \S\ref{sec:sfrs}. The recomputed numbers for the dust-corrected \ha/\oii~G10 SFR ratios (lower left panel of Fig.~\ref{sfratio}, given in \S4.3) using $A_{\rm Wuyts}$ or $A_{\rm gas} = 1.4A_{\rm stars}$ give instead: 1.60$\pm$0.74 and 0.91$\pm$0.46 for the low mass bin and 2.60$\pm$1.30 and 3.00$\pm$1.50 for the high mass bin respectively. The latter being only marginally inconsistent (at $\lsim$1.5$\sigma$) with the prediction of the G10 correction.

\subsection{Stacking of non-detections}
To further investigate the spectra where we do not detect \ha,~we stack these spectra in three bins in stellar mass (tabulated in Table \ref{stackparams}). We first wish to test if we are able to make a significant detection from these data. Thus, to ensure that the final spectrum is not dominated by one or two detections close to our nominal threshold of 5.0$\sigma$, we only stack spectra with a \sn$<$4.5. All spectra in a given mass bin are first shifted to a reference wavelength of 13500 \textrm{\AA} (corresponding to a redshift, \z$_{\rm ref}$ = 1.057), which is the centre of the G141 grism wavelength coverage. For each stellar mass bin, each spectrum is scaled in flux relative to the luminosity distance using \z$_{\rm ref}$. The spectra are then summed and divided by the number of spectra. The stacked spectra are analyzed as described in \S3 to obtain the line luminosity and its total \sn~(Eq. \ref{sneqn}). In all the stacked spectra, we are able to detect a significant \ha~line. The aperture used for measuring the flux is sufficiently wide that the residual wavelength uncertainties (\S\ref{sect:lineveri}) do not affect the summed flux measured within it. The \sn~of these three average spectra are higher than our detection threshold meaning that we could use these as extra data points in each of the mass bins when calculating the SSFR. Table \ref{stackparams} gives the \sn~and number of stacked spectra for each mass bin.

\begin{table}
\begin{tabular}{ l l l}
\hline 
Bin width (log(M$_{*}$/M$_{\odot}$) & N$_{spec}$ & \sn\\
\hline
8.6 - 9.5 & 40 & 23.5\\
9.5 - 10.5 & 43 & 42.2\\
10.5 - 11.5 & 6 & 4.9\\
\hline
\end{tabular}
\caption{For \sn$_{\rm Ha}<4.5$ detections, the sizes of the mass bins and the number of galaxies they contain, the number of stacked spectra, and the \sn~of the stacked spectrum in each mass bin.}
\label{stackparams}
\end{table}

\section{The SSFR-mass Relation at z$\sim$1}
With these measurements in hand, we can now turn our attention to the SSFR-mass relation at z$=$1. The SSFR is simply computed by dividing the \ha~SFR measurement by the galaxy's stellar mass. Firstly, we consider the impact of the non-detections on the average relation measured. Next we compare with other (shallower) works using \ha~at similar redshifts, and finally we compare our measurements with the recent results from state-of-the art numerical simulations \citep[EAGLE\footnote{the public database is accessible via \url{http://icc.dur.ac.uk/Eagle/}},][]{Schaye2015:mnras521,2015MNRAS.450.4486F}.

\begin{table}
\begin{tabular}{ l l l }
\hline 
Bin width (log(M$_{*}$/M$_{\odot}$) & N$_{gal}$ & Median SSFR (yr$^{-1}$)\\
\hline
\ha~detections only & & \\
8.6 - 9.5 & 28 & -8.60 $\pm$ 0.09 \\
9.5 - 10.5 & 16 & -9.13 $\pm$ 0.11 \\
10.5 - 11.5 & 12 & -9.34 $\pm$ 0.14 \\
\hline
Stack & & \\ 
8.6 - 9.5 & 60 & -8.93 $\pm$ 0.09 \\
9.5 - 10.5 & 39 & -9.58 $\pm$ 0.10 \\
10.5 - 11.5 & 12 & -9.77 $\pm$ 0.29 \\
\hline
\end{tabular}
\caption{
The sizes of the mass bins and the number of galaxies they contain together with the median SSFR for each mass bin}
\label{ssfrparams}
\end{table}

The average relation is found by taking the median SSFR in bins of mass, as tabulated in Table \ref{ssfrparams}. In order to account for the impact of non-detections, we construct a weighted stack by combining the stacked spectra described above with the individual detections at higher S/N. This is our main result and shown as the cyan points in Fig.~\ref{ssfr}. The scatter on these measurements is estimated by bootstrap resampling the data in each bin, treating all data points as detections whether formally detections or limits. This should provide a not unreasonable estimate of the uncertainty on the average relation.

\begin{figure*}
\noindent \begin{centering}
\scalebox{1.5}{\includegraphics[width=9.0cm,height=8cm]{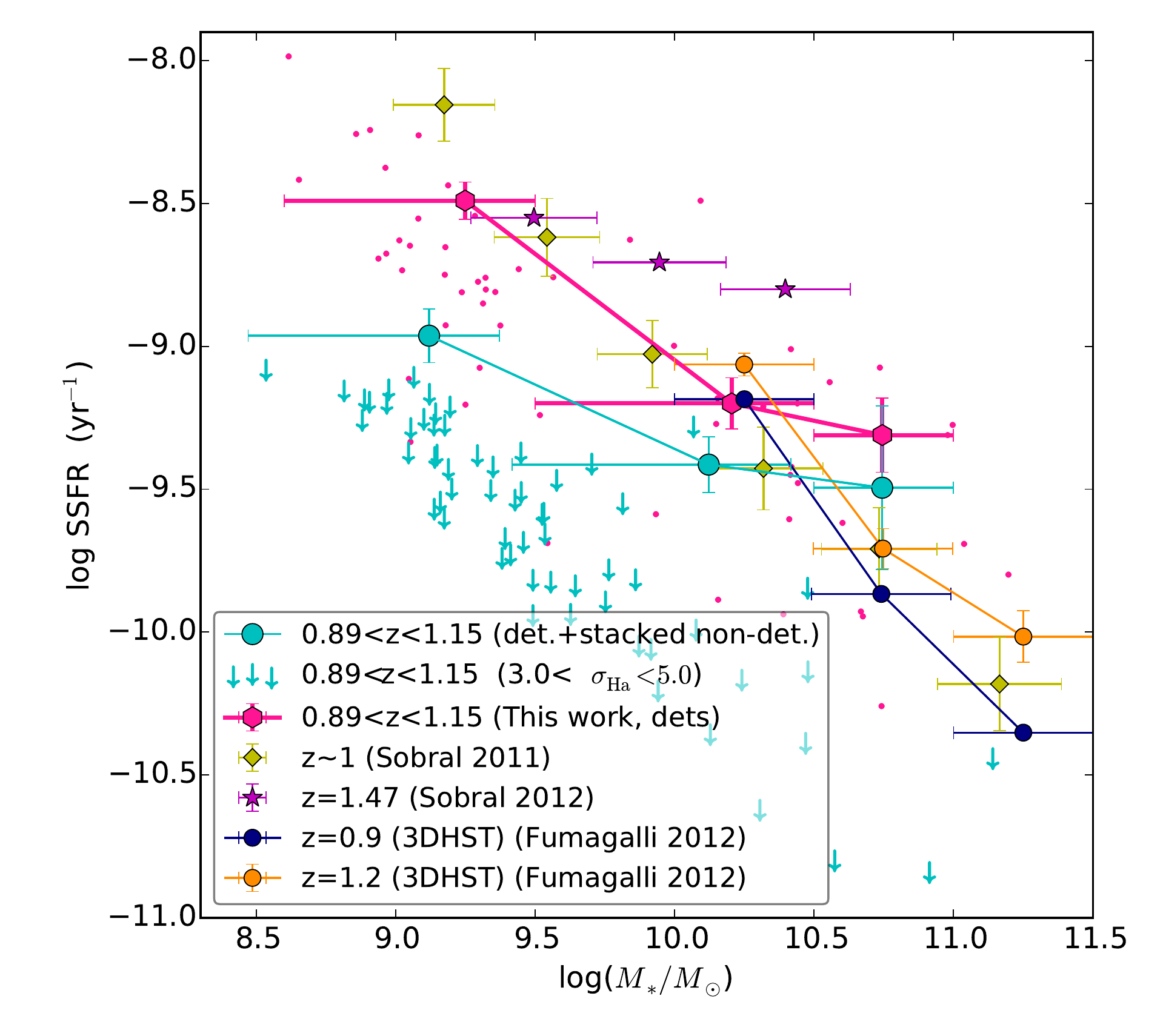}}
\par\end{centering}
\caption[Caption]{SSFR-mass relation where the SSFR has been measured with H$\alpha$. (For consistency, the 3D-HST points from Fumagalli$+$ have been corrected for \nii~contamination in the same way as described for our data.) We compare our median SSFR (detection) measurements (pink circles; individual galaxies shown as pink points) and stacked results including non-detections (cyan circles and upper limits) to the measurements from \cite{2012ApJ...757L..22F} (navy and orange circles), \citealt{2011MNRAS.411..675S} (yellow diamonds) and  \cite{2012MNRAS.420.1926S} (magenta stars). Our data probe to much lower ($\sim$1 dex) SSFRs than existing spectroscopic samples at similar redshifts and our stacked measurement indicates the flattening of the SSFR data missed by shallower (incomplete) samples. See text for discussion and also next figure.}
\label{ssfr}
\end{figure*}

A number of works have recently attempted to make either narrowband or spectroscopic measurements of \ha~SFRs at these redshifts and their published measurements are also shown in Fig.~\ref{ssfr}. The HiZELS (The High-Redshift Emission Line Survey) study by Sobral et al.\ (2009, 2011) used deep NIR narrow-band imaging in J, H and K bands with the Wide Field Camera at UKIRT. Sobral et al. (2012), built on the samples detected in individual bands by using a matched \ha$+$\oii~dual narrow-band survey to obtain a wide-field sample at z$\sim$1.5. In order to obtain deeper data over a wider area, Fumagalli et al.\ (2012) used \tdhst~space-based spectroscopic data to study a mass-selected sample of galaxies between 0.8 $<$ z$<$ 1.2. The samples closest to our redshift range are the \cite{2012ApJ...757L..22F} and \cite{2011MNRAS.411..675S} at \z$\sim$0.9 and \z=1 respectively. In the highest mass bins our results are consistent. However, when moving towards lower stellar mass, it is clear that our results (even for the detections only) lie systematically below other works. This is due to the greater depth of our survey, and incompleteness in the other surveys. \citet{2011MNRAS.411..675S} note that their sample selection is restricted to a fixed SFR limit of $>$3 \Msunyr. At the lowest masses probed, low SSFR galaxies fall below their selection limit, biasing their median SSFRs upwards. They found that this bias becomes significant at masses below $\sim$10$^{10}M_{\odot}$. Hence, our sample gives a better SSFR measurement for the low mass (low SFR) galaxies. Similarly, \citet{2012ApJ...757L..22F}, who also used \textit{HST} WFC3 grism data, only probed massive galaxies ($>$10$^{10}M_{\odot}$) at a detection limit corresponding to SFR = 2.8 \Msunyr.\footnote{Applying a cut of SFR$>$3\Msunyr to our data brings our measurements closer to the other two measurements in the lowest mass bin, but doesn't entirely explain the difference. The point is moved by $\sim$0.1dex. In the higher mass bins the points are essentially unchanged.}

One point worth emphasising in our study is the greater depth achievable with the \textit{same WFC3 grism data} by pre-selecting galaxies based on the ROLES' \oii~spectroscopic redshifts and by stacking those objects not formally detected individually. This can also be seen by comparison with the recent public release of the \tdhst~\citep{2015arXiv151002106M} spectroscopic catalogue. See comparison in Appendix~\ref{sec:cf3dhst}.

\begin{table}
\begin{tabular}{ l l  }
\hline 
Sample & SSFR--mass slope, $\alpha$ \\
\hline
This work, detections & -0.50 $\pm$0.04\\
This work, stack & -0.47 $\pm$0.04\\
Sobral et al.\ 2011 (z$\sim$1) & -1.00 $\pm$0.07\\
EAGLE (z$\sim$1) & -0.14 $\pm$0.05\\
Stripe 82 (z$\sim$0.1), \logmass<10 & -0.08 $\pm$0.01\\
\hline
\end{tabular}
\caption{The slopes of the SSFR--mass relation for some of the results shown in Figs.~\ref{ssfr} and \ref{average_ssfr}. Representative examples are taken at z$\sim$1, except the Stripe 82 work taken as a local comparison. See text for discussion.}
\label{table:slopes}
\end{table}

It is clear from this figure that the steep slope of the SSFR-stellar mass relation depends greatly on the depth of the data used. Given the difficulty in detecting weak \ha~sources, this is not surprising, and it emphasises the point that complete samples are vital.  To illustrate this, faint-end slope power law fits (SSFR$ \propto M_\star^\alpha$) for representative sample are given in Table~\ref{table:slopes}. As can be seen, the literature work, such as Sobral et al.\ (2011), gives  $\alpha \approx$-1.0. We measure flatter slopes of $\alpha =$ -0.50 $\pm$0.04 for our detections only, and $\alpha =$-0.47 $\pm$0.04 for our stack. Fortunately the greater depth of our ROLES' OII detections allows us to obtain the average \ha~of a complete sample. Our conservative $A_{\rm gas} = A_{\rm stars}$ dust scaling probably leads to a steeper slope than in reality as it likely underestimates the size of the correction (and hence the SSFR) at the highest stellar masses. Applying the alternative dust corrections of $A_{\rm Wuyts}$ and $A_{\rm gas}$ = 1.4 $A_{\rm stars}$ negligibly changes the results, giving values of $\alpha$ = -0.46$\pm$0.05 and -0.46$\pm$0.04 respectively. 

The flatter slope measured with our deeper data has important consequences since, as we discuss below, the flatter relation implies a more uniform average star formation history for galaxies of low and intermediate mass.

One additional point which merits comment is the offset between the ROLES \oii~relation and the \ha~relation measured in this work. The current work uses a subset (from GOODS-S) of the galaxies used in the Gilbank et al.\ 2010 \oii~result. In order to compare directly, the dashed line in Fig.~\ref{average_ssfr} uses the same \oii~measurements for exactly the same galaxies used in the \ha~weighted sum (our preferred method, large cyan circles). It can be seen that this lies intermediate between the Gilbank et al.\ 2010 \oii~result and the current \ha~result. We showed in \S\ref{sec:sfrs} that the two indicators agree on average, when measured for \ha~detections (i.e. the subset of galaxies represented by the magenta hexagons in Fig.~\ref{average_ssfr}). However, it is perhaps not too surprising that the ratio of \ha/\oii~might be slightly higher than the measured value ($\sim$unity) if lower \ha~SFRs (non-detections) are omitted. It must be remembered that the error bars do not include the uncertainty/scatter in the transformation from \oii~to \ha. Thus it is reassuring that the apparent systematic offset between the \oii~SFRs for all the ROLES WFC3 galaxies (dashed) line lies only $\approx$0.1dex between both the Gilbank et al.\ 2010 \oii~SSFR--mass relation, and our current \ha~relation. A thorough investigation of this point will require deeper spectroscopy.

A final note is that the absolute calibration of \ha~SFR is dependent on the K98 calibration which is likely uncertain at the 0.2 dex level. The independent calibration of \citet{Chang2015:apjs8} would lower the normalisation of the observed measurements by 0.2 dex.

\subsection{Comparison with the EAGLE simulations}

In order to place our results in a theoretical context, we compare to the Evolution and Assembly of GaLaxies and their Environments (EAGLE) simulations. The simulations follow the formation of galaxies (and black holes) in a cosmologically representative (100\,Mpc$^3$) volume \citep{Crain2015:mnras1937,Schaye2015:mnras521}. The simulations use advanced smoothed particle hydrodynamics (SPH) and state-of-the-art subgrid models to capture the unresolved physics and include reionisation, cooling, metal enrichment and energy input from stellar feedback. Black hole growth and feedback from active galactic nuclei are also included. A complete description of the code and physical parameters used can be found in \citet{Schaye2015:mnras521} and \citet{Crain2015:mnras1937}. These simulations have reproduced  many of the observed properties of the Universe with unprecedented fidelity (for example, the evolution of the stellar mass function, the evolution of specific star formation rates; see \citealt{2015MNRAS.450.4486F}), providing a powerful tool to study galaxy formation and evolution \citep{Schaye2015:mnras521,Trayford2015:mnras2879,Schaller2015:mnras2277,Lagos2015:mnras3815,Rahmati2015:mnras2034}. The evolution of the cosmic star formation history was not used to directly set parameters for the model. The models were calibrated to yield the z $=$ 0.1 galaxy stellar mass function and central black hole masses as a function of galaxy stellar mass. The prediction of a reasonable cosmic SFH appears to be a consequence of reproducing the observed sizes of present-day galaxies (as well as the local galaxy stellar mass function). As shown in \citet{Crain2015:mnras1937}, simulations that reproduce the observed sizes also match the observed star formation history of the Universe well. The star formation histories of galaxies are extensively discussed in \citet{2015MNRAS.450.4486F}, and we use this as the basis for comparison with our observational 
measurements. We use the halo and galaxy catalogues available in the EAGLE database at \url{http://www.eaglesim.org/database.php} \citep{McAlpine2016:AstronomyandComputing72}, and focus on galaxies in the
Recal-L025N0752 calculation. The particle mass of this calculation is $2.26\times10^{5}M_{\odot}$ for gas particles and $1.21\times10^{6}M_{\odot}$ for dark matter particles. The particle mass is a factor 8 smaller that than of the larger volume
Ref-L100N1504 calculation, allowing us to study the star formation histories of 
galaxies as small as $10^8M_{\odot}$ in stellar mass.

\begin{figure*}
\noindent \begin{centering}
\includegraphics[width=18.0cm]{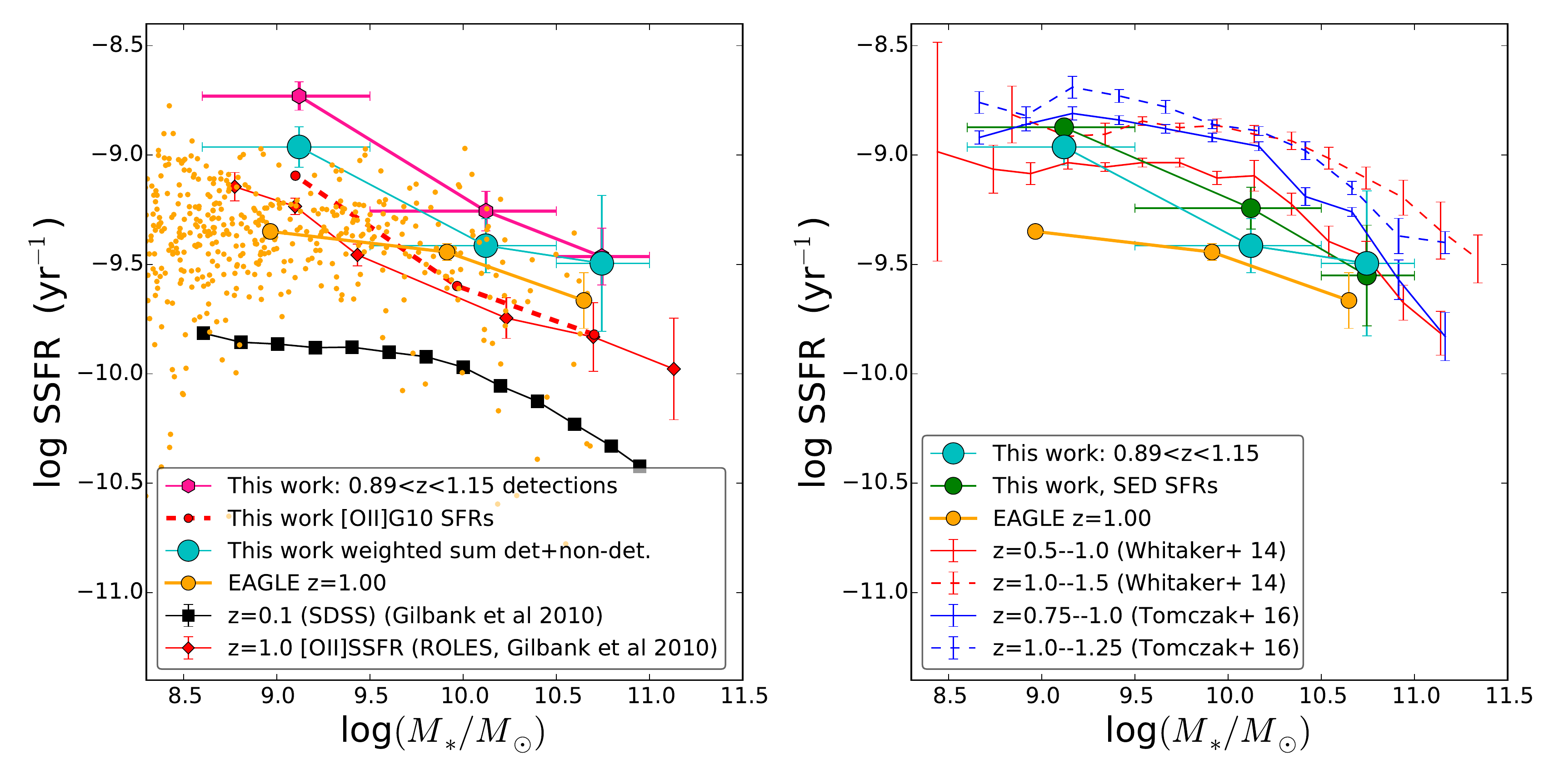}
\par\end{centering}
\caption[]{SSFR-mass relation where the SSFR has been measured with H$\alpha$ for this work at \z$\sim$1 (symbols as for previous figure); \oii ~from \citealt{2010aMNRAS.405.2594G} (red circles); and \ha~from stripe 82 at z$\sim$0.1 (left panel).  Results from the EAGLE simulation \recal~are shown as orange points, with the median shown as larger connected circles. In addition, the right panel shows results from two recent photometric works. Refer to text for discussion.}
\label{average_ssfr}
\end{figure*}

In Fig.~\ref{average_ssfr} we show the simulated galaxies as orange points, with the average SSFR--stellar mass in the simulation shown as an orange line. Error bars denote the scatter on the mean. Again, our main result from the stacking analysis is shown in cyan. At low masses, the simulation predicts a relation that is almost flat below a mass of $10^{10.5}$. Since the SSFR can be interpreted as the inverse of the galaxy formation timescale, a flat distribution of specific star formation rates implies that galaxies of all stellar masses have similar star formation histories. 
This is a natural outcome of the simulations since dark matter haloes have similar growth rates regardless of stellar mass \citep{Fakhouri2010:mnras2267,Correa2015:mnras1514}. The flatness of relation
is therefore a generic feature of galaxy formation models \citep{2012MNRAS.422.2816B}: relations 
that rise or decline strongly with stellar mass require indicate that an additional 
physical scale is in involved in galaxy formation. Such a scale is indeed seen in the SSFR of more massive galaxies as star formation is strongly suppressed by feedback 
from black hole growth in galaxies more massive than $10^{10.5}M_{\odot}$ \citep[][Bower et al. 2016, MNRAS submitted]{2006MNRAS.365...11C,2006MNRAS.370..645B,Dubois2015:mnras1502} 
The normalisation of the SSFR--stellar mass relation reflects the steepness of the stellar-mass halo-mass relation of galaxies and its evolution \citep{Moster2010:apj903,Behroozi2010:apj379}.

In low mass galaxies in EAGLE, star formation is regulated by outflows from supernova feedback (implemented by stochastically heating particles in the vicinity of newly-formed stars). The mass loading of the outflow is larger in smaller galaxies, because of the binding energy of the halo, which leads to the relatively flat faint-end slope of the galaxy mass function \citep[][Bower et al. 2016, MNRAS submitted]{Crain2015:mnras1937}.

Comparison between the observed results and the average SSFR--mass relation derived from the stacking analysis demonstrate the critical importance of accounting for selection effects. Without accounting for the upper-limits, the data would naively suggest a steeply rising SSFR at lower masses. Such a result would be in stark contrast to the simulation results and would imply the existence of an additional physical process that is not modelled in the EAGLE simulation. However, the H$\alpha$ emission of most low mass galaxies is not detectable, and such upper limits must be carefully taken into account.  
Through our stacking analysis, we are able to better estimate the average specific star formation rate as a function of stellar mass and the average SSFRs of low mass galaxies drop considerably, bringing the observations more closely in line with the simulation predictions. However, although we measure a significantly flatter low mass slope than other works, our result is still significantly steeper than that predicted by EAGLE ($\alpha =$ -0.14 $\pm$0.05, c.f. Table \ref{table:slopes}).

Fig~\ref{average_ssfr} also shows the SSFR--mass relation for local galaxies, determined from SDSS Stripe 82 \ha~data \citep{2010aMNRAS.405.2594G}, together with results from the EAGLE simulations.
At low reshift the simulations and observations agree well (see \citealt{2015MNRAS.450.4486F} for
discussion of the differences between the higher resolution Recal-L025N0752, shown here,
and the reference simulation Ref-L100N1504). It is a remarkable success of the simulation 
that the rate of evolution of the relation is so well described, showing that the simulation correctly captures the decline in the cosmic star formation rates of galaxies
that results from the slowing accretion of dark matter haloes at the present day. In order to strengthen our results, deeper \ha~spectrosocpy aimed at  detecting individually the lowest mass objects would both allow us to measure the intrinsic scatter in the SSFR--mass relation, and confirm whether or not our current sample still shows some residual incompleteness, implying at an even flatter low mass slope.

\subsection{Comparison with photometric studies}
The majority of work on the low mass end of the SSFR--mass relation has used photometric information, due to the difficulty in obtaining spectroscopy of sufficient depth. The obvious disadvantages of using only photometry are the potential inaccuracy of photometric redshifts, particularly given the faint magnitudes necessarily associated with these low mass objects; the systematic differences between SFR indicators using photometric techniques, particularly their sensitivity to longer timescales than that probed by \ha~(see, e.g., fig.~3 of \citealt{2010bMNRAS.405.2419G}); and the degeneracy between these two (the covariance between redshift and SFR is rarely properly accounted for in these fits). However, they obviously allow much larger samples to be constructed and it is instructive to compare a couple of recent results with our spectroscopy and with each other. Fig.~\ref{average_ssfr} (right panel) shows the ROLES and EAGLE data together with results from \citet{2014ApJ...795..104W} and \citet{Tomczak2016:apj118}\footnote{We use their ``SFRsf" sample as being closest to our selection. Using ``SFRall" lowers the SSFR marginally (by including lower SSFR systems), but still within the intra-study difference.}. In each case, two redshift bins contain the ROLES redshift range, so the two closest bins are shown for each sample. In both cases, there is good agreement between our measurements in our lowest and highest mass bins, however, our intermediate mass bin (\ms$\sim$10) is lower than both other surveys. This leads to a steeper slope in our data than either of the others. Interestingly both surveys show good agreement in overall normalisation at the low mass end (where the statistics should be best), with the \citet{Tomczak2016:apj118} result being marginally higher. All these works show higher overall normalisation at the low mass end  than the EAGLE prediction. To see if these differences could be due to the SFR indicator used, we recalculate the median SSFR--mass relation using our SED-fit SFRs, fitted at the \ha~spectroscopic redshift (or \oii~where \ha~was undetected). This gives the green line shown in Fig.~\ref{average_ssfr}. All of our points agree with our \ha~measurements, within the uncertainties, but the intermediate mass bin has moved upwards to be more consistent with the other photometric SSFR--mass relations. This hints that the differences between studies may, at least partly, be due to the different SFR indicators used. However, although the most discrepant data point is now closer to the other works, this leads to a formally marginally steeper slope ($\alpha=-0.37\pm0.04$). Aside from differences in techniques (spectroscopic vs photometric measurements), another important consideration is cosmic variance. Although our field has some overlap with these other works, this is actually a small fraction of the individual photometric samples. \citet{2014ApJ...795..104W} covers at total area of 900 arcmin$^2$ from several fields in CANDELS, \citet{Tomczak2016:apj118} covers a total area of $\approx$400 arcmin$^2$. In comparison, our total surveyed area in \ha~is $\approx$100 arcmin$^2$ in GOODS-S. Our redshift range covered is comparable to the bin sizes used in \citet{Tomczak2016:apj118} 0.89$<$z$\leq$1.15. So, the ratio in volumes is approximately the ratio in survey area. See \citet{Tomczak2016:apj118} for some discussion of the likely impact of cosmic variance on these works. Given the agreement with our higher and lower mass bins, it is unlikely that cosmic variance is responsible for the difference in intermediate masses and more likely reflects a real difference in slope between the measurements. Resolving this will again require larger deep spectroscopic samples. 

\section{Conclusions}
We have studied a sample of low-mass galaxies ($M_{*}$$\sim$10$^{8.5}$$M_{\odot}$) taken from the ROLES survey at \z$\sim$1 with the aim of determining their SFRs using a more direct SFR indicator, \ha.  These systems were analyzed using NIR slitless spectroscopic data from \textit{HST}. We measured the \ha~emission line luminosities and converted these into SFRs.
\begin{itemize}
\item We have shown, by comparison of our SED-inferred dust extinction with Balmer decrement corrected SFRs, that the Calzetti relation for estimating gas phase extinction ($A_{\rm gas} = 2.27 A_{\rm stars}$) overestimates the extinction in \ha, in line with other recent works. We adopt a fiducial relation of $A_{\rm gas} = 1.0 A_{\rm stars}$ as a lower limit to the SFR throughout but discuss alternative choices motivated by other recent results.
\item In this study, we compare our H$\alpha$ SFRs with the \oii~SFRs from ROLES. We find that the ratio of the \ha~and \oii~SFRs is consistent, within the broad uncertainties, with the \oii G10~mass-dependent empirical relation calibrated at \z$\sim$0.1. The fact that these two SFRs are comparable implies that the technique for measuring SFRs using ground-based spectroscopy, which is currently more efficient than obtaining \ha~SFRs from space, is accurate. 
\item We measure a flatter low mass power law slope (-0.47 $\pm$0.04) to the z$\sim$1 SSFR--mass relation than found by other (shallower) \ha-selected samples ($\approx$-1), although still somewhat steeper than that predicted by the EAGLE simulation (-0.14 $\pm$0.05), hinting at possible missing physics not modelled by EAGLE or remaining incompleteness for our \ha~data.
\end{itemize}

Deeper \textit{HST} slitless spectroscopy would allow us to attempt to directly detect the \ha~emitters contributing to our stack, confirming these stacking results and enabling a measurement of the intrinsic scatter in the SSFR--mass relation in this critical low mass regime. These data would also provide a more direct estimate of completeness, addressing whether there is room in the observational results for an even flatter low mass slope.

\section*{Acknowledgements}
We thank the anonymous referee for useful comments which improved this work. We thank Tom Jarrett and Andrew Hopkins for comments on an earlier form of this work, as RR's MSc thesis. This work is based on observations taken by the 3D-HST Treasury Program (GO 12177 and 12328) with the NASA/ESA HST, which is operated by the Association of Universities for Research in Astronomy, Inc., under NASA contract NAS5-26555. RR, DGG, SLB and RES acknowledge financial support from the National Research Foundation (NRF) of South Africa, RR additionally acknowledges support from the National Astrophysics and Space Science Programme (NASSP), and the University of Cape Town. MLB would like to acknowledge support from an NSERC Discovery grant. RGB acknowledges the support of STFC consolidated grant ST/L00075X/1. 

The data used in this paper may be obtained on request from the corresponding author.

\bibliographystyle{mnras}
\bibliography{ms}

\appendix
\section{Flux repeatability test}
In our sample there are galaxies that have multiple spectra from repeated \textit{HST} observations in  overlappping fields. These spectra are used to test the repeatability of our flux measurements, the \sn~and wavelength accuracy.

 Originally, the spectrum with the lowest $\sigma$ was chosen for our catalogue (as is commonly done when combining catalogues). We show why this leads to a bias in the H$\alpha$ measurements, and our alternate approach to negate this bias. 
 
 The repeatability of our flux measurements was tested by comparing the ratio of  the observed errors to the expected errors for each. This ratio was computed as follows,
 \begin{equation}
\frac{F(H\alpha)_{used}-F(H\alpha)_{rep}}{\sqrt{\sigma_{used}^{2}+\sigma_{rep}^{2}}}
 \end{equation}
where $F(H\alpha)_{used}-F(H\alpha)_{rep}$ is the observed error obtained from the difference between the line flux used in the catalogue (F(H$\alpha$)$_{used}$) and the flux from repeat observations (F(H$\alpha$)$_{rep}$). The expected error is the sum in quadrature of the errors on the flux measurement used in the catalogue ($\sigma_{used}$) and from repeat observations ($\sigma_{rep}$). If these errors have been calculated properly, a normal distribution, with a mean of zero, is expected. It was found that there is an offset in the mean ($\mu\sim$1.5). This offset is because we initially chose the lowest $\sigma$ spectrum to use in the catalogue which creates a bias. We corrected this bias by replacing the lowest $\sigma$ spectrum for a galaxy with multiple spectra with another randomly selected spectra to use in the catalogue.  This reduces the mean offset to 0.68. In addition, calculating the dispersion of this second sample gives $\sigma=$1.0, showing that our flux uncertainties are repeatable. 

\section{S/N Repeatability}
\label{sect:snrep}
In \S \ref{sec:spec} we chose a detection threshold of 5$\sigma$ (S/N$\geq$5) based on visually inspecting spectra where there were obvious bright emission lines. In this section, we test whether this threshold is reasonable and if it really corresponds to a 5$\sigma$ detection. As mentioned before, choosing a threshold is a trade off between purity and completeness. We can test our threshold by determining the reproducibility of the emission lines for galaxies that have multiple spectra. Our criteria for whether a detection is real or not is that the line should be found in the majority of the spectra (e.g. if a galaxy has four spectra, the line should be seen in three out of the four spectra) to be considered reproduced. The best way to test this is to pick a spectrum where the line has a very high \sn~because it is more likely to be a real detection. An example of a galaxy that has four spectra, with a high \sn, is shown in Fig. \ref{highsn}. A line is seen in all the spectra meaning that the line is reproduced. 

\begin{figure*}
\centering
\includegraphics[width=18cm]{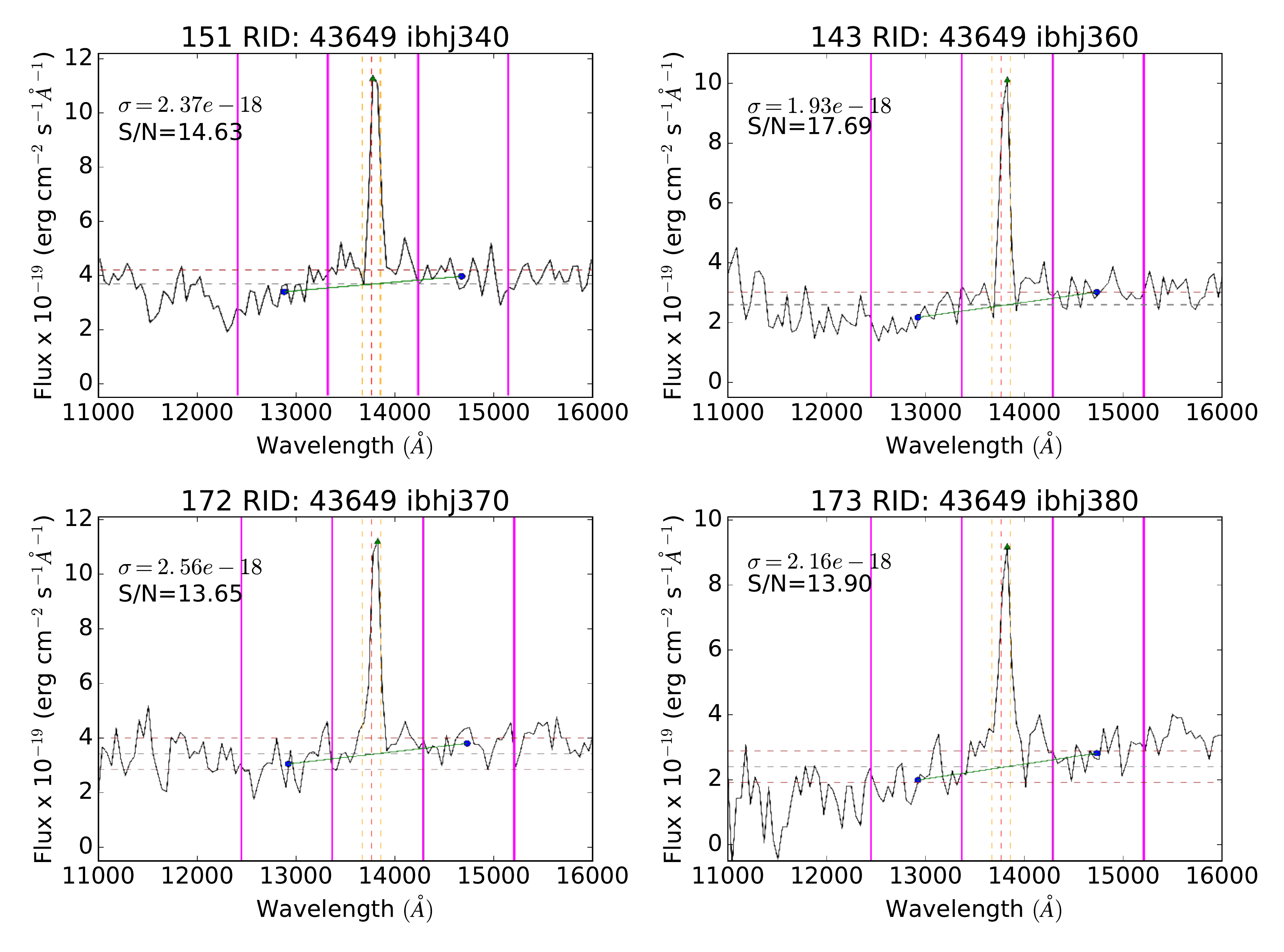}
\caption{Example of a galaxy's spectrum with a high \sn~emission line in four pointings. The emission line is seen in all pointings meaning that it is reproduced. Lines and symbols have the same meaning as in Fig. \ref{lie}}
\label{highsn}
\end{figure*}

\begin{figure}
\noindent \begin{centering}
\includegraphics[width=0.5\textwidth]{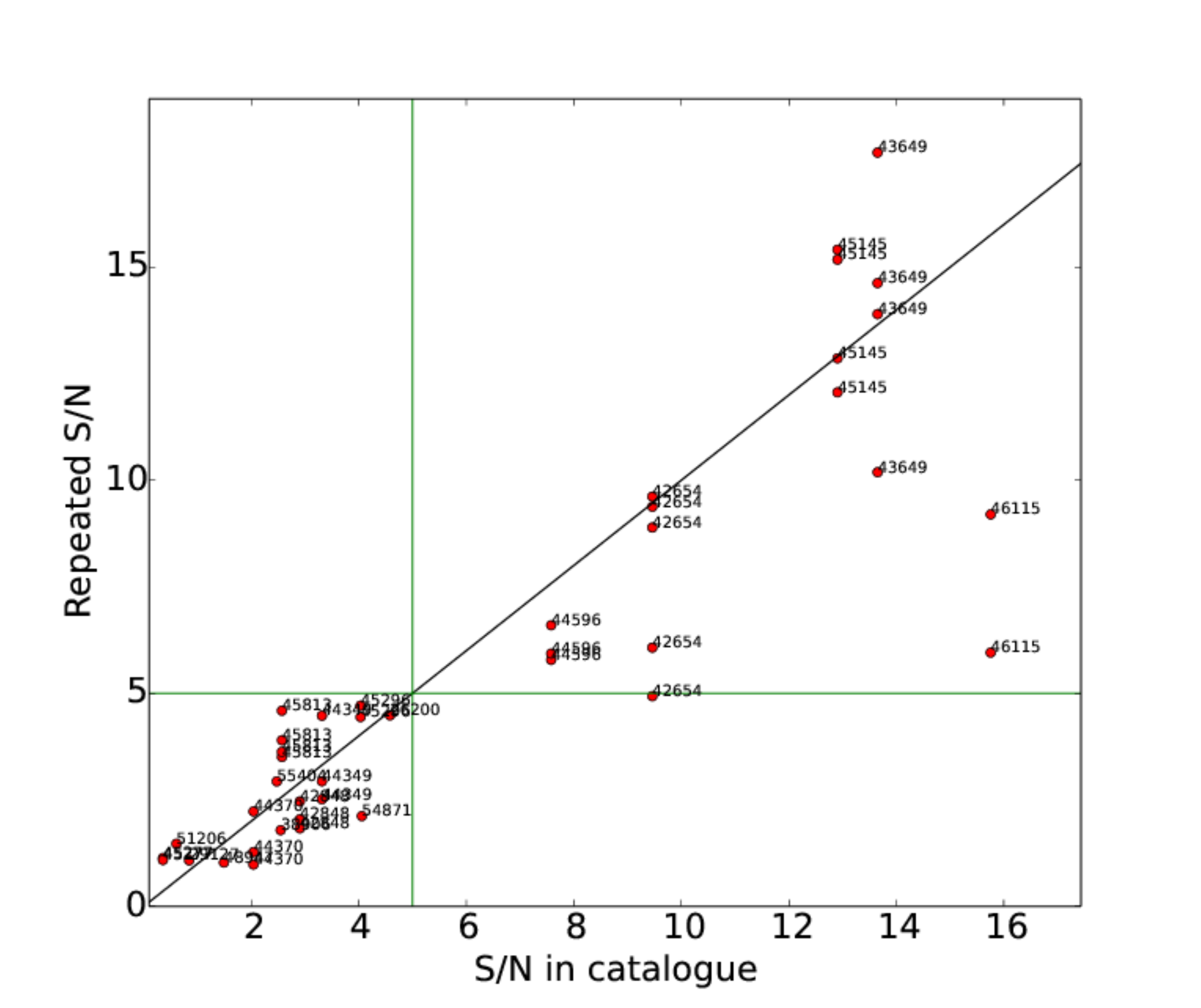}
\par\end{centering}
\caption{\small{\sn~from repeat spectra against S/N used in catalogue. Green lines indicate the \sn$=$5 threshold.}}
\label{snr}
\end{figure}

\begin{figure}
\noindent \begin{centering}
\includegraphics[width=0.5\textwidth]{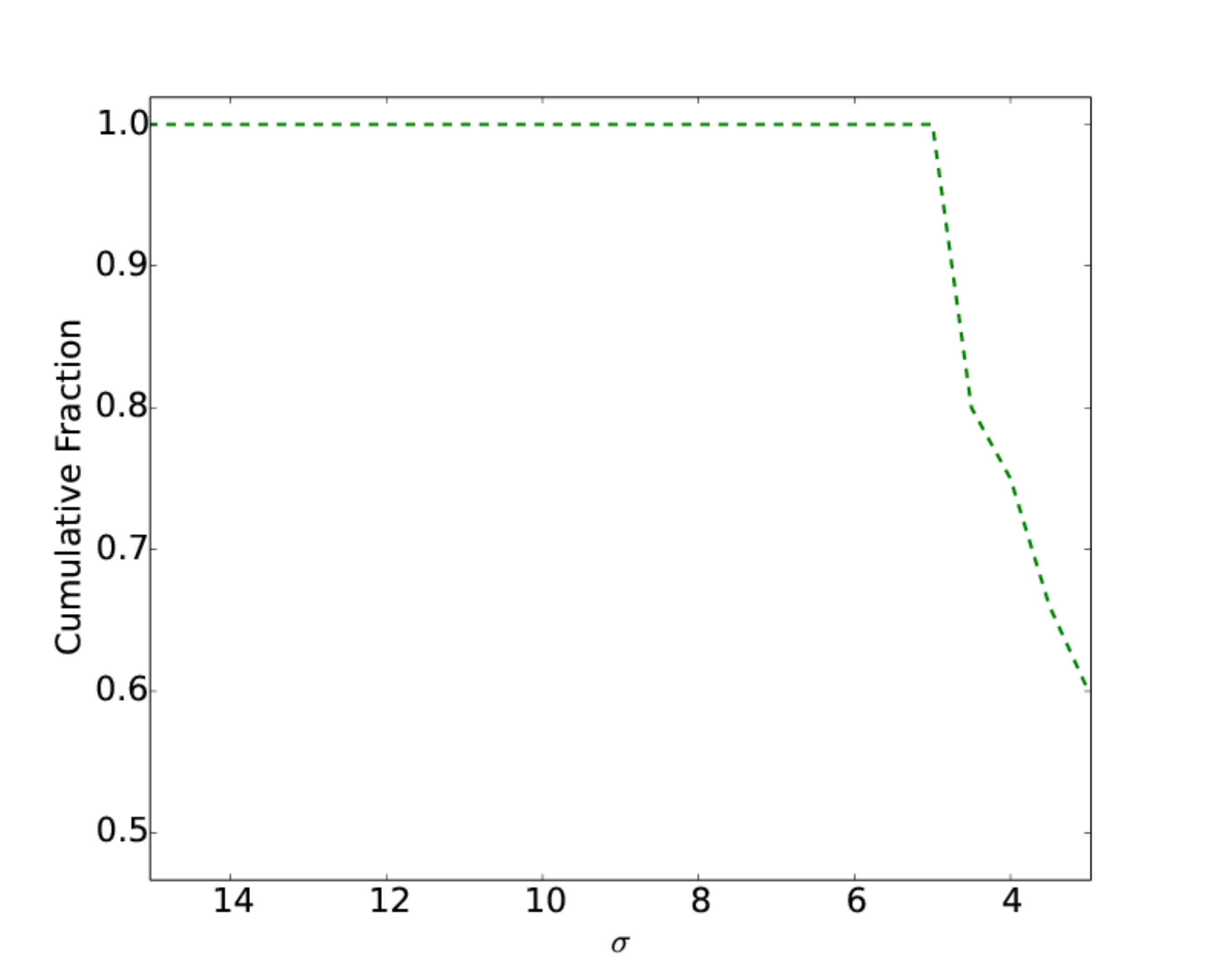}
\par\end{centering}
\caption{\small{Recovered fraction of galaxies from our sample of galaxies with multiple spectra.}}
\label{recfr}
\end{figure}

In Fig. \ref{snr} we plot the S/N used in our catalogue against the S/N from repeat spectra. Moving from right to left (high to low \sn) in Fig. \ref{snr},  we look at the points in the vertical direction because these correspond to the same galaxy and check to see if they lie above our chosen threshold. If they do, we count them as recovered. However, the chosen threshold has some uncertainty on it so when deciding if a spectrum is recovered, we also count those points that lie fairly close, within $\delta$\sn$=$0.3, to the chosen threshold as recovered. We only go down to \sn$=$3 to see how many galaxies we can recover because we do not believe any detections below \sn$=$3. The fraction of recovered galaxies is plotted in Fig. \ref{recfr}.
This plot shows that for our sample of galaxies that have repeat spectra, we recover 60-100$\%$ of galaxies  between \sn$=$3 and 5. Based on Fig.~\ref{recfr}, if we look at the \sn~distribution in our catalogue (see Fig. \ref{hist}) and apply a \sn$=$4.5 threshold, for example, we get 18 detections out of which 80$\%$ are real and 20$\%$ are spurious. Our initial threshold of \sn$=$5 thus gives us a pure sample but we are missing some real detections.

\section{Comparison with \tdhst}
\label{sec:cf3dhst}
With the recent public release \citep{2015arXiv151002106M} of \tdhst~reductions and derived products (redshifts, emission line fluxes, etc.) of the spectroscopic data we have used in this work, it is possible to compare their results with the independent measurements made by the current work. Fig.~\ref{cf3dhstz} compares the redshifts of the two works. In the left panel, the two redshifts are plotted against one another, and in the right panel the redshift difference is shown as a function of apparent $K$-band magnitude, with a histogram showing the number of objects as a function of the residual. In these plots, objects with a ROLES' redshift but not a corresponding \tdhst~redshift are plotted at -1.0 on the vertical axis in the left plot and in the region above 1.5 in the vertical axis on the right panel. From these, it is clear that there is overall good agreement where a redshift is measured in both works. Where \ha~is detected in our analysis, only two out of 50 galaxies show disagreement at the $\Delta z>0.01$ level (which is $\sim3\sigma$ level set by the resolution of the grism spectra) with \tdhst~redshifts. Recall that ROLES is based on \oii~detections in deeper optical spectroscopy (in terms of SFR probed at z$\sim$1). Where an emission line was detected in ROLES, its probablity of being \oii~was compared with the probability of being another line using photometric redshift (photo-z) PDFs. This resulted in a probability of being \oii~versus that of being another common line visible in the ROLES' wavelength window, wt(\oii).\footnote{See \citet{2010bMNRAS.405.2419G} for more details. Briefly, almost every single emission line detection was more likely to be \oii~than another, lower redshift line. A weighting was calculated based on the ratio of photoz PDFs and if the ratio was $>$0.9 in favour of \oii~it was set to unity. Conversely, if the probability was <0.1, its value was set to 0. Everything in between was kept as the formal ratio of probabilities of other redshifts and these measurements received fractional weighting in our analysis.} For \tdhst, redshifts are estimated from full spectral fits to the grism spectroscopy, combined with photo-z information from an updated photometric dataset \citep{2014ApJS..214...24S}. So, in principle, \tdhst~redshifts should be more secure than that used by the ROLES parent sample. However, the average ROLES galaxy is a $K$-faint galaxy with little continuum in the WFC3 grism data, and likely relatively noisy photometry from which to derive a photo-z. Thus, it is not surprising that a significant fraction (7/57 for \ha~detections, and 140/281 in the whole ROLES sample) have no secure grism redshift from \tdhst. The results of this comparison are tabulated in Table~\ref{tab:cf3dhst}.

\begin{figure*}
\noindent \begin{centering}
\hspace*{-0.6cm}{\includegraphics[width=8.cm]{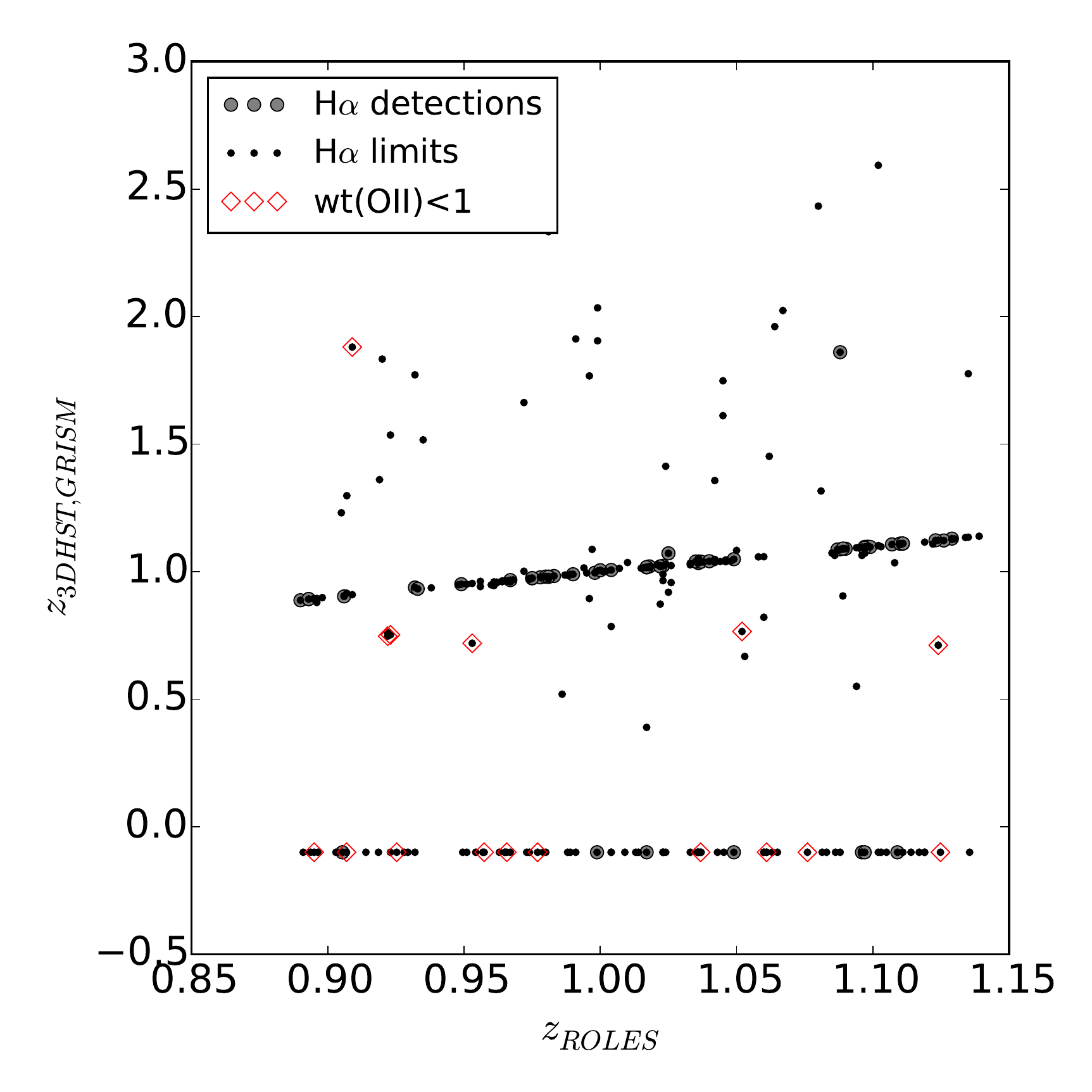}\includegraphics[width=8.0cm]{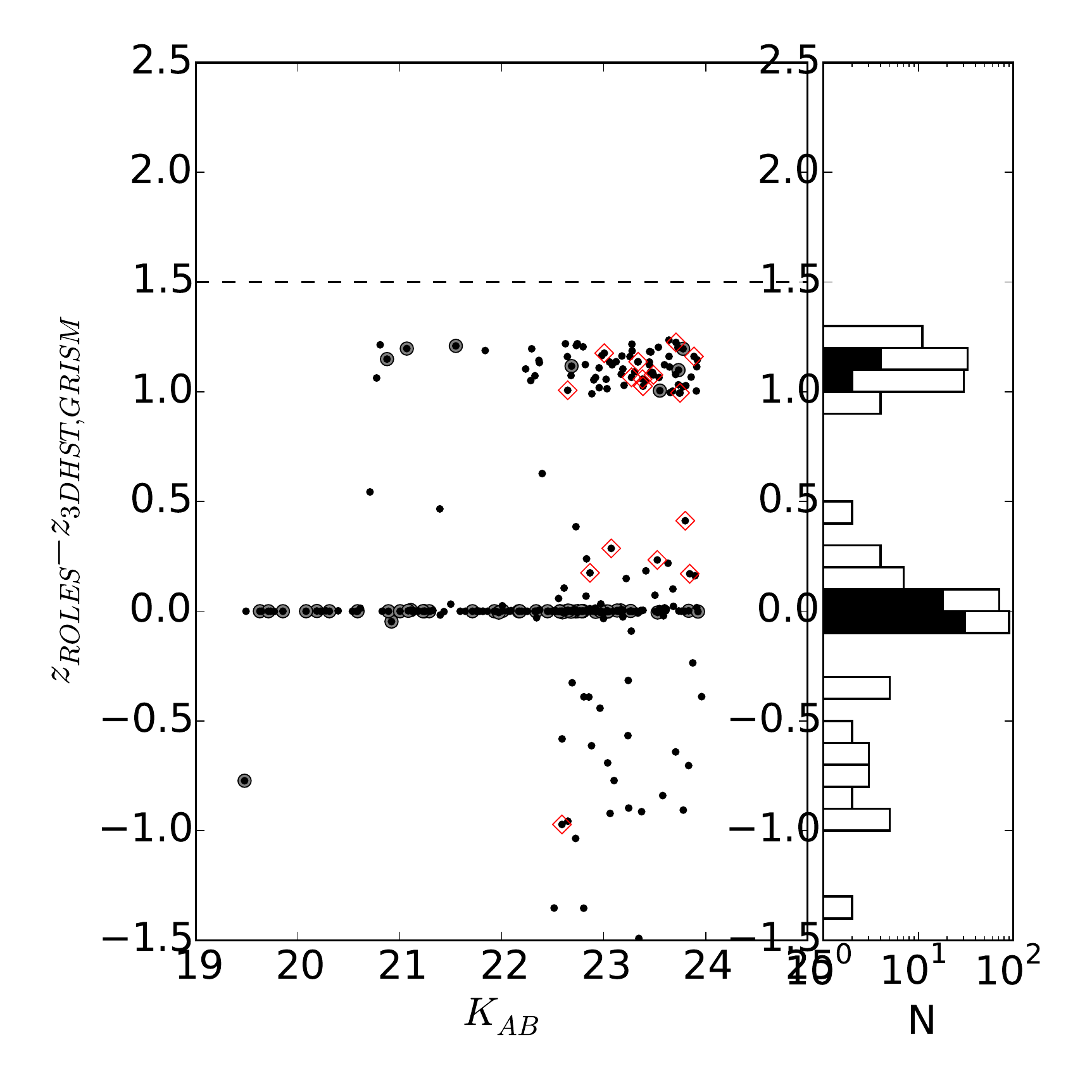}}
\par\end{centering}
\caption{Comparison of our redshifts with 3DHST. Left panel: our redshifts vs those of 3D-HST. Larger filled symbols denote $\sigma\ge$5 detections in our data; open red diamonds show lower confidence ROLES redshifts with wt(\oii)$<$1 (see text for discussion). Right panel: redshift difference as a function of apparent $K$-band magnitude. Symbols are same as in left panel. Objects with no corresponding redshift in 3D-HST are shown at $\Delta$ z$>$1.5 (and z$<$0 in left panel). Histograms show the number of objects in the whole sample (open histogram) and those with ROLES \ha~detections (filled historgam). Note the log scale. }
\label{cf3dhstz}
\end{figure*}

\begin{figure}
\noindent \begin{centering}
\hspace*{-0.6cm}{\includegraphics[width=8.cm]{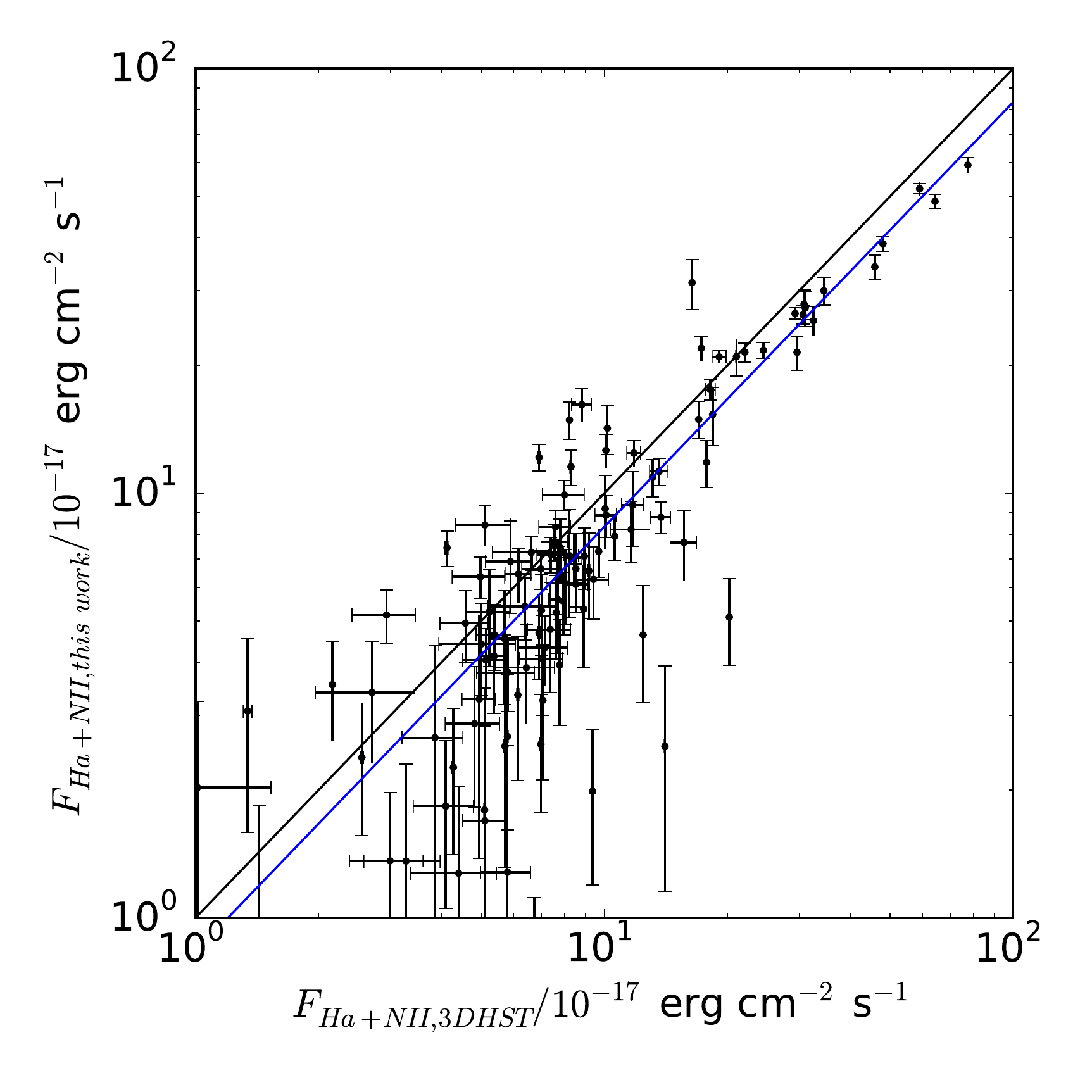}}
\par\end{centering}
\caption{Comparison of our \ha$+$\nii~fluxes with those measured by \tdhst. This shows that our simple aperture measurement compares favourably to the technique used by  which attempts a more sophisticated fitting of the continuum, plus a renormalisation based on the multiwavelength imaging data avaialable. The black line shows the one-to-one relation. It appears that the \tdhst~fluxes are higher than ours by about 20\% (indicated by blue line). See text for discussion.}
\label{cf3dhstF}
\end{figure}

\begin{table}
\caption{Summary of comparison with 3D-HST shown in Fig.~\ref{cf3dhstz}. See text for discussion.}
\centering
\begin{tabular}{llll}
\hline
Sample & N in & N with  & N with no \tdhst\\
 & sample  & $\Delta z<0.01$ & \ha~detection\\
\hline
ROLES \ha~detections only & 57 & 48 & 7 \\
ROLES \ha~detections $+$ limits & 281 & 78 & 140\\
\hline
\end{tabular}
\label{tab:cf3dhst}
\end{table}

Fig.~\ref{cf3dhstF} shows a comparison with \tdhst~\ha$+$\nii~fluxes \citep{2015arXiv151002106M}. In this comparison, neither catalogue is corrected for the contribution of \nii. Our simple aperture measurement shows reasonable agreement with the  sophisticated technique used by \citet{2015arXiv151002106M}, however we see an overall trend that our fluxes lie systematically around 20\% fainter. \citet{2015arXiv151002106M} used model fitting of the continuum (compared with our local linear fit) plus a renormalisation of the absolute flux based on extensive multiwavelength photometry. So, although we use the same \textit{HST} spectroscopy (albeit with a different extraction and measurement method), the overall calibration appears systematically different. Shifting our calibration to agree for this would only raise our \ha~luminosities, and hence SFRs by 0.08 dex.

\subsection{Completeness}

The spectroscopic completeness of the parent ROLES sample used in this work (\oii-selected) was presented in \citet{2010bMNRAS.405.2419G} (see in particular fig.~12). For the GOODS-S field it was estimated to be $\gsim$80\% independent of magnitude, down to the survey limit of $K$=24.0. The 3D-HST data allows us to reassess this with the benefit of deeper data and higher precision photo-zs. However, it must be borne in mind that photometric redshifts are not well-tested at these faint magnitudes. In \citet{2010bMNRAS.405.2419G}, we showed how we targeted almost every galaxy in our $K$-band magnitude range $(22.5 < K \le 24.0$, independent of photo-z, and only used the (FIREWORKS) photo-z PDFs to break degeneracies between most likely emission lines at other redshifts outside our target window ($0.89<z\le1.15$). With these caveats in mind, we can proceed to examine the galaxies which \tdhst~photometric data places within our desired stellar mass and redshift ranges and see what fraction we a) placed slits on, b) obtained \oii~detections. 

Fig.~\ref{fig:cpltness} shows the completeness estimated by comparison with \tdhst~photometric redshifts. The left panel shows the ROLES-observed targets as a function of $K$-band magnitude. The magnitudes (and masses) are taken from \tdhst~photometry, in order to be consistent with the photometric redshifts used. Several caveats must be noted. Since ROLES used $K$-band selection ($22.5 < K_{AB} \le 24.0$), differences between $K$ photometry in \tdhst~vs FIREWORKS can lead to objects being lost from this plot. Indeed, at the faintest magnitudes there is significant scatter ($\sim$0.3 mag). The left panel shows all the slits we placed (where `slits' specifically refers to ROLES slits and does not include the ESO brighter sample, since we do not have the targeting information for that) for all objects (regardless of \tdhst~photometric redshift); those within the redshift window (equivalent to spectroscopic/targeting completeness); and those which resulted in an emission line consistent with \oii~(equivalent to redshift success rate). The right panel shows the equivalent as a function of stellar mass. Again, an important caveat is that the stellar masses here all come from \tdhst~(except where noted) and so there is some degeneracy between fitting the stellar mass and photometric redshift to the same photometry. Typically the covariance between the parameters is ignored, and the best-fit photoz is simply assumed to be correct.  In order to assess the possible impact of this, the redshift success rate is shown as a function of stellar mass calculated from our own SED-fitting at the spectroscopic redshift, and \tdhst's stellar masses at their (mostly photometric) redshifts. 

The overall targeting completeness is somewhat lower than calculated in  \citet{2010bMNRAS.405.2419G} which was approximately 80\% independent of stellar mass. This is likely due to a combination of the reasons mentioned above (difference in $K$-band photometry between \tdhst~and FIREWORKS; improved precision of photozs, and inclusion of grism redshifts, versus FIREWORKS broader photo-z PDFs); but also might reflect the limitations of the different photometric redshifts at these faint magnitudes. Indeed, an instructive exercise is to plot independent photometric redshift estimates from different codes from the same data at these faint magnitudes and the scatter is generally surprisingly large  (M. Franx, priv. comm.)! So, the completenesses shown in the plot should be regarded as lower limits.

For the higher mass, ESO public spectroscopic sample \citep{2008A&A...478...83V}, pertinent details of the selection were discussed in \citet{2010bMNRAS.405.2419G} and we briefly summarise the salient points here. The survey was $z$-band selected but otherwise comparable to ROLES in terms of spectroscopic depth, and the 1D spectra were processed for emission line measurements in the same way as for ROLES. Although the follow-up comprised specific sub-samples, such as X-ray selected targets, we rejected the small number of objects in our redshift window showing both \oii~and X-ray emission (as possible AGN) as well as likely AGN from \textit{Spitzer} colours. Thus we found the spectroscopic (targeting) completeness to be $\sim$40\% in each $K$ magnitude bin except for the brightest bin ($K$<22.75) which is about 70\% complete. 

Finally, it is worth noting that our highest mass SSFR measurements are in good agreement with literature values, and so our measurements of the slope of the SSFR relation are not limited by uncertainties in the high mass (ESO public spectroscopic) sample.

\begin{figure*}
\noindent \begin{centering}
\hspace*{-0.6cm}{\includegraphics[width=18.cm]{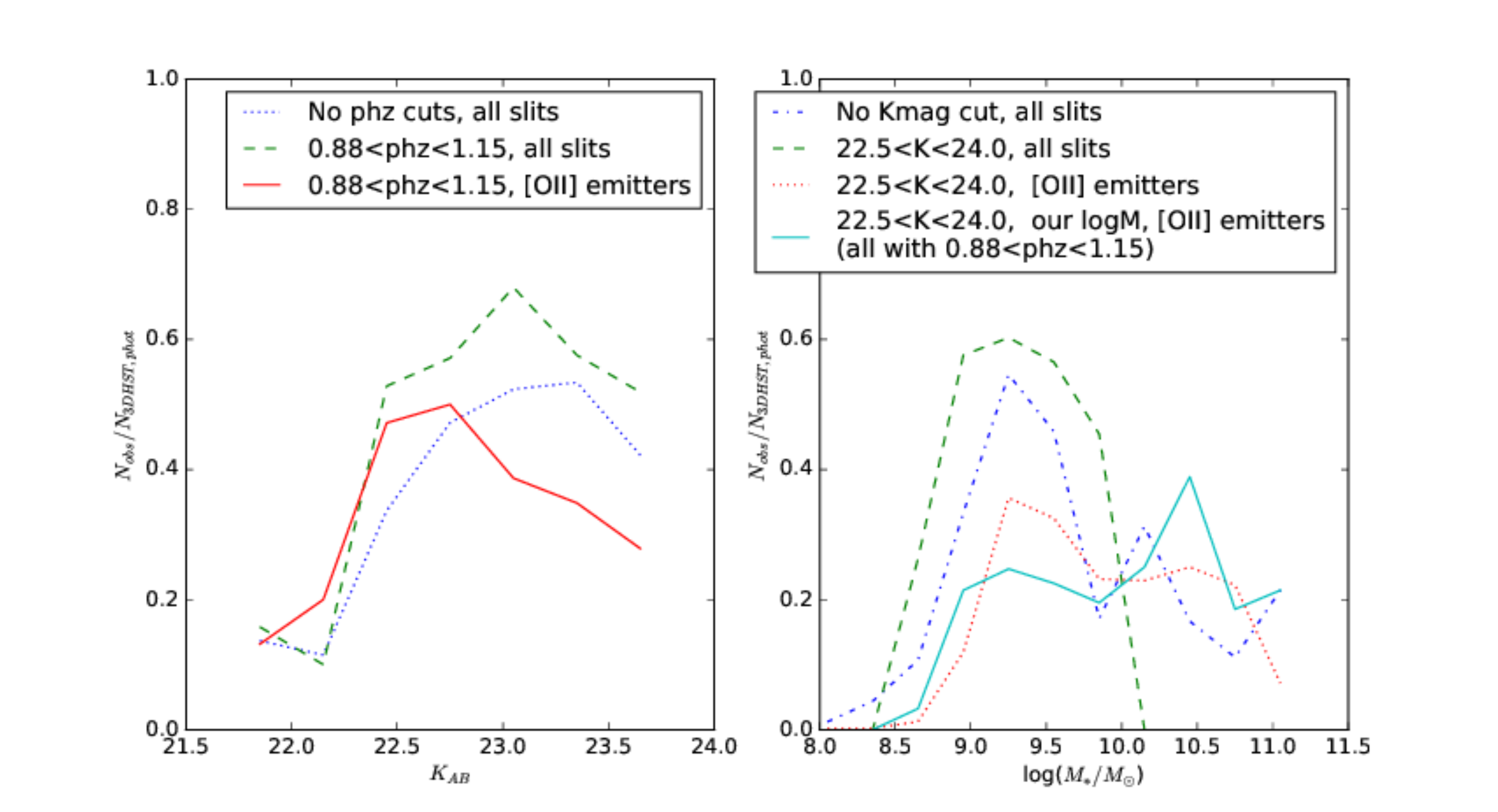}}
\par\end{centering}
\caption{Left panel: test of completeness as a function of apparent $K$-band magnitude from \tdhst~catalogues. Results are shown for all objects regardless of photo-z; for those with \tdhst~photozs in our redshift window (``spectroscopic/targetting completeness"); and for those which resulted in \oii~detections (``redshift success rate"). Right panel: similar tests as a function of stellar mass. In all cases except solid cyan line, masses come from \tdhst. See text for discussion.}
\label{fig:cpltness}
\end{figure*}

\bsp	
\label{lastpage}
\end{document}